\documentclass[aps,reprint,twocolumn,10pt,amsmath,superscriptaddress,amssymb,longbibliography]{revtex4-2}
\usepackage{epsfig,epstopdf,dcolumn,amsbsy,bm,color,CJK}
\usepackage{amsfonts,appendix,graphicx,algorithm,enumerate,makeidx,braket,verbatim}
\usepackage[dvipsnames]{xcolor}
\usepackage[driverfallback=dvipdfmx, colorlinks, linkcolor=blue, citecolor=blue, urlcolor=blue]{hyperref}
\usepackage[all]{hypcap}

\begin{document}

\title{Two-Dimensional Quantum Droplets in a Mixture of Bose-Einstein Condensates}

\author{Yue Zhang}
\affiliation{State Key Laboratory of Quantum Optics Technologies and Devices, Institute of Opto-electronics, \\
\makebox[0.8\linewidth][c]{Collaborative Innovation Center of Extreme Optics, Shanxi University, Taiyuan, Shanxi 030006, China}}

\author{Liangchao Chen}
\email[Contact author: ]{chenlchao87@sxu.edu.cn}
\affiliation{State Key Laboratory of Quantum Optics Technologies and Devices, Institute of Opto-electronics, \\
\makebox[0.8\linewidth][c]{Collaborative Innovation Center of Extreme Optics, Shanxi University, Taiyuan, Shanxi 030006, China}}
\affiliation{Hefei National Laboratory, Hefei, Anhui 230088, China}

\author{Wei Han}
\affiliation{State Key Laboratory of Quantum Optics Technologies and Devices, Institute of Opto-electronics, \\
\makebox[0.8\linewidth][c]{Collaborative Innovation Center of Extreme Optics, Shanxi University, Taiyuan, Shanxi 030006, China}}
\affiliation{Hefei National Laboratory, Hefei, Anhui 230088, China}

\author{Linshan Gong}
\affiliation{State Key Laboratory of Quantum Optics Technologies and Devices, Institute of Opto-electronics, \\
\makebox[0.8\linewidth][c]{Collaborative Innovation Center of Extreme Optics, Shanxi University, Taiyuan, Shanxi 030006, China}}

\author{Yazhou Wang}
\affiliation{State Key Laboratory of Quantum Optics Technologies and Devices, Institute of Opto-electronics, \\
\makebox[0.8\linewidth][c]{Collaborative Innovation Center of Extreme Optics, Shanxi University, Taiyuan, Shanxi 030006, China}}

\author{Zekui Wang}
\affiliation{State Key Laboratory of Quantum Optics Technologies and Devices, Institute of Opto-electronics, \\
\makebox[0.8\linewidth][c]{Collaborative Innovation Center of Extreme Optics, Shanxi University, Taiyuan, Shanxi 030006, China}}

\author{Xinhao Liang}
\affiliation{State Key Laboratory of Quantum Optics Technologies and Devices, Institute of Opto-electronics, \\
\makebox[0.8\linewidth][c]{Collaborative Innovation Center of Extreme Optics, Shanxi University, Taiyuan, Shanxi 030006, China}}

\author{Wenjie Sui}
\affiliation{State Key Laboratory of Quantum Optics Technologies and Devices, Institute of Opto-electronics, \\
\makebox[0.8\linewidth][c]{Collaborative Innovation Center of Extreme Optics, Shanxi University, Taiyuan, Shanxi 030006, China}}

\author{Pengjun Wang}
\affiliation{State Key Laboratory of Quantum Optics Technologies and Devices, Institute of Opto-electronics, \\
\makebox[0.8\linewidth][c]{Collaborative Innovation Center of Extreme Optics, Shanxi University, Taiyuan, Shanxi 030006, China}}
\affiliation{Hefei National Laboratory, Hefei, Anhui 230088, China}

\author{Lianghui Huang}
\affiliation{State Key Laboratory of Quantum Optics Technologies and Devices, Institute of Opto-electronics, \\
\makebox[0.8\linewidth][c]{Collaborative Innovation Center of Extreme Optics, Shanxi University, Taiyuan, Shanxi 030006, China}}
\affiliation{Hefei National Laboratory, Hefei, Anhui 230088, China}

\author{Zengming Meng}
\affiliation{State Key Laboratory of Quantum Optics Technologies and Devices, Institute of Opto-electronics, \\
\makebox[0.8\linewidth][c]{Collaborative Innovation Center of Extreme Optics, Shanxi University, Taiyuan, Shanxi 030006, China}}
\affiliation{Hefei National Laboratory, Hefei, Anhui 230088, China}

\author{Zhuxiong Ye}
\affiliation{State Key Laboratory of Quantum Optics Technologies and Devices, Institute of Opto-electronics, \\
\makebox[0.8\linewidth][c]{Collaborative Innovation Center of Extreme Optics, Shanxi University, Taiyuan, Shanxi 030006, China}}

\author{Jing Zhang}
\email[Contact author: ]{jzhang74@sxu.edu.cn}
\affiliation{State Key Laboratory of Quantum Optics Technologies and Devices, Institute of Opto-electronics, \\
\makebox[0.8\linewidth][c]{Collaborative Innovation Center of Extreme Optics, Shanxi University, Taiyuan, Shanxi 030006, China}}
\affiliation{Hefei National Laboratory, Hefei, Anhui 230088, China}

\date{\today }

\begin{abstract}
Quantum droplets represent a striking manifestation of beyond-mean-field physics, with properties that depend sensitively on dimensionality. Although quantum droplets have been experimentally realized in one and three dimensions, their realization in two dimensions has remained elusive. Here, we observe the formation of a two-dimensional (2D) quantum droplet in a quasi-2D geometry by transferring the single-component Bose-Einstein condensates (BECs) into a spin mixture in the Lee-Huang-Yang (LHY) regime, where the LHY effect evidently deviates from its three-dimensional form. We find that, in our scheme, 2D quantum droplets are more stable and exhibit longer lifetimes than previously reported quantum droplets. By instead transferring the initial BECs into another single-spin state with a negative scattering length, we generate 2D Townes solitons. A comparison between the two types of self-bound states shows that, whereas 2D Townes solitons are stable only near a critical atom number determined by the scattering length, 2D quantum droplets remain well-defined and self-bound over a much broader parameter range. These realizations of quantum droplets and Townes solitons in two dimensions provide new insights into self-bound states and open new avenues for exploring nonlinear dynamics in quantum liquids.
\end{abstract}

\maketitle

Quantum droplets are remarkable self-bound states in ultracold atomic systems that can maintain their shape in free space and exhibit liquid-like behavior~\cite{PRL.115.155302, Science359.301, PRL.120.235301, PRL.120.135301, PRR.1.033155, PRR.3.033247}. This property arises from a delicate balance between attractive mean-field interactions and the repulsive beyond-mean-field Lee-Huang-Yang (LHY) correction. As a direct manifestation of beyond-mean-field quantum fluctuations, they offer a powerful platform to explore quantum many-body physics~\cite{RPP.84.012403, Nature574.382, PRR.4.043018, PRA.102.043304, PRA.102.053308} and nonlinear dynamics~\cite{PRL.122.090401, PRL.126.244101, PRL.123.160405}, attracting significant theoretical and experimental interest. The LHY correction manifests in different forms depending on the dimension, thereby changing the phase diagram of quantum droplets~\cite{PRL.117.100401, PRA.98.051604, FP.16.32201, PRL.122.105302, PRL.133.083401, PRR.3.012027, PRA.98.051603}. To date, quantum droplets have been experimentally realized in three-dimensional (3D) and effectively one-dimensional (1D) geometries in several different platforms, including dipolar Bose-Einstein condensates (BECs)~\cite{Nature539.259, PRX.6.041039} and both homonuclear~\cite{Science359.301, PRL.120.135301, PRL.120.235301} and heteronuclear Bose mixtures~\cite{PRR.3.033247, PRR.1.033155}.  Numerous investigations have focused on revealing distinctive properties of quantum droplets, such as surface tension effects~\cite{PRL.122.090401}, collective oscillation behaviors~\cite{PRL.126.230404, PRL.121.173403}, and capillary instability~\cite{PRL.134.093401}.

Matter-wave solitons are another kind of localized self-bound state of ultracold atoms that can maintain their shape during propagation~\cite{Science296.1290, Nature417.150, NP10.918, PRL.123.123602, PR.899.1, CP.7.163, PRL.135.263404}. Generally, stable solitons can be found in 1D systems, and stricter conditions will be required for their formation in higher-dimensional propagation spaces~\cite{EPJST.225.2507, NRP.1.185, PRL.90.040403}. For instance, in 1D optical waveguides, by adjusting the atomic interaction of single-component BECs to be weakly attractive, a delicate balance between wave-packet dispersion and interatomic attraction can be achieved, leading to the formation of stable bright solitons~\cite{Science296.1290, Nature417.150, NP10.918, PRL.123.123602}. When the interactions are tuned to be repulsive, dark or gray solitons can be realized~\cite{PRL.83.5198, Science287.5450, NP.4.496, JPA.43.213001}. Bright soliton trains or soliton gases can also emerge from matter waves extended in 1D waveguides. The complete growth process of modulation instability, which eventually leads to the formation of soliton trains, can be observed in experiments~\cite{Science356.422, PRL.128.013201}. In particular, in contrast to both three- and one-dimensional systems, two-dimensional (2D) systems possess scale invariance, which allows for the existence of Townes solitons~\cite{PRL.13.479, PRL.125.250401, PRL.127.023604, PRL.127.023603}. However, these states exist only at a discrete value of the product of the atom number and the interaction strength, rendering them unstable against atom loss.

%%%%%%%%%%%%%***Fig. 1***%%%%%%%%%%%%%%%%%%
 \begin{figure}[htbp]
\centerline{\includegraphics[scale=1.0]{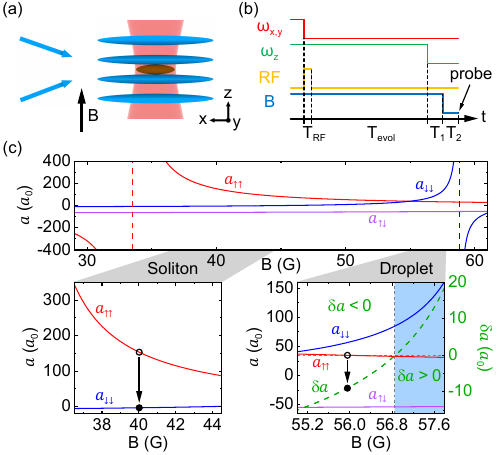}}
\caption{(a) Schematic illustration of the quasi-2D geometry. The ultracold atoms are compressed into a quasi-2D geometry with axial confinement frequency $\omega_z$ by an accordion optical lattice formed by the interference of two blue-detuned $532\,\mathrm{nm}$ laser beams. A red-detuned $1064\,\mathrm{nm}$ laser beam propagating vertically provides weak confinement with frequency $\omega_{x,y}$ in the xy-plane. (b) Experimental sequence for the preparation of quantum droplets and Townes solitons. (c) Magnetic-field dependence of the scattering lengths near the Feshbach resonance. Quantum droplets are formed in a two-spin mixture with $\delta a < 0$, whereas Townes solitons are created in the single $\lvert\downarrow\rangle$ state with $a_{\downarrow\downarrow} < 0$. The black arrows indicate the RF-pulse-driven transitions between the spin states.}
\label{fig1}
\end{figure}

%%%%%%%%%%%%%****Fig. 1***%%%%%%%%%%%%%%%%%%

The different stabilization mechanisms of bright solitons and quantum droplets give rise to distinct dimensional dependences, which can be understood from the size-dependent energy functional~\cite{PRL.115.155302, PRL.117.100401, PRL.120.135301}. In a quasi-1D system with tunable transverse confinement, for ultracold atoms (single- or two-component) with only attractive mean-field interactions, the energy functional exhibits a minimum at a finite wave-packet size, supporting stable bright solitons~\cite{Science296.1290, PRL.123.123602}. The LHY effect creates a new minimum in the energy functional, leading to the formation of quantum droplets~\cite{PRL.122.105302, PRA.98.013631}. By adjusting the atom number or interaction strength, the relative depths of the two minima can be tuned, thereby enabling the crossover between droplets and solitons~\cite{PRL.120.135301, PRR.8.023090}. In three dimensions, the energy functional of attractive mean-field interactions develops a singularity with an infinitesimal wave-packet volume, which precludes the existence of solitons~\cite{PT.72.46}. With the LHY effect, the energy functional has a minimum, corresponding to the formation of quantum droplets in 3D free space~\cite{PRR.3.033247, PT.72.46}. In a quasi-2D system with tunable axial confinement, for attractive mean-field interaction at the critical atom number, the energy functional reduces to a flat profile, giving rise to marginally stable Townes solitons~\cite{PRL.127.023604, PRL.127.023603}. Deviating from the critical atom number, Townes solitons will collapse or dissociate. Theoretical studies indicate that the LHY effect induces a minimum in the energy functional of the quasi-2D system, thereby enabling droplet formation over a broader range of atom numbers~\cite{PRL.117.100401, PRR.3.012027}. Moreover, on the relatively flat mean-field energy profile in 2D, the LHY correction can more readily induce a local minimum at a larger cloud size, potentially allowing droplets much larger than those observed in previous experiments. Nevertheless, experimental realizations and studies of 2D droplets remain scarce.

Here, we create quantum droplets in a quasi-2D geometry with tight axial confinement, where the LHY effect evidently deviates from its 3D form, leading to emergent properties of the 2D quantum droplets. Our experiment begins with $^{39}$K BECs confined in a crossed optical dipole trap, with atoms initially prepared in the $\lvert\uparrow\rangle\equiv\lvert 1,-1\rangle$ spin state. Subsequently, we employ an accordion optical lattice formed by two interfering $532\,\mathrm{nm}$ laser beams to compress the BECs into a single horizontal layer of the lattice (see Fig.~\hyperref[fig1]{1(a)})~\cite{CPB.31.103401}. For the accordion optical lattice with a $5\,\mu\mathrm{m}$ spacing, the vertical trapping frequency (z-direction) is set to $3.67\,\mathrm{kHz}$, which is sufficient to counteract gravity and compress BECs into a quasi-2D regime. Thereafter, a vertically propagating $1064\,\mathrm{nm}$ laser beam is ramped on to provide weak in-plane confinement with a trapping frequency of approximately $42\,\mathrm{Hz}$, followed by the removal of the crossed optical dipole trap.

The experimental sequence for preparing quantum droplets and Townes solitons from quasi-2D BECs is presented in Fig.~\hyperref[fig1]{1(b)}. Quantum droplets and Townes solitons are produced by magnetically tuning the atomic interactions via a Feshbach resonance. Thus, we first adjust and stabilize the magnetic field strength to the target point, then turn off the vertical $1064\,\mathrm{nm}$ beam and immediately apply a radio-frequency (RF) $\pi /2$-pulse ($40\,\mu\mathrm{s}$) for droplets or an RF $\pi$-pulse for solitons to transfer the atoms to the spin state $\lvert\downarrow\rangle \equiv \lvert 1,0 \rangle$. After allowing the ultracold atoms to evolve for a certain period in the quasi-2D geometry, we turn off the accordion optical lattice for detection. As the internal energy of the ultracold atoms is mainly released in the vertical direction, the atomic cloud expands rapidly in this direction, whereas the horizontal expansion is markedly slower~\cite{SM}\nocite{PRL.84.2551, PRA.106.013320, Nonlinearity21.R139, OL.39.3429}. Consequently, the in-plane density profile remains nearly unchanged and closely resembles the in situ distribution. After about $2\,\mathrm{ms}$ of expansion at the initially set scattering length, by which time the cloud is sufficiently diluted, we switch off the Feshbach magnetic field and allow an additional $2\,\mathrm{ms}$ of time-of-flight expansion at the low-magnetic-field scattering length. The droplets and solitons are then detected by absorption imaging along the vertical direction.

The atomic scattering lengths are tuned to produce quantum droplets and Townes solitons using the Feshbach resonance of the $\lvert\downarrow\rangle$ state at $58.86\,\mathrm{G}$. The corresponding magnetic-field regions are shown in Fig.~\hyperref[fig1]{1(c)}. In the two-spin mixture for producing quantum droplets, the effective mean-field interaction near the optimal population ratio $N_\uparrow/N_\downarrow =\sqrt{a_{\downarrow\downarrow}/a_{\uparrow\uparrow}}$ is characterized by the effective scattering length $\delta a = a_{\uparrow\downarrow} + \sqrt{a_{\uparrow\uparrow} a_{\downarrow\downarrow}}$, where $a_{\uparrow\uparrow}$ and $a_{\downarrow\downarrow}$ are the intra-spin scattering lengths, and $a_{\uparrow\downarrow}$ is the inter-spin scattering length~\cite{PRL.115.155302, Science359.301}. Using the Feshbach resonance, we can tune $\delta a$ between positive and negative values. In an appropriate regime with $\delta a<0$, the relative contribution of the LHY correction becomes significant, enabling the formation of quantum droplets. For ultracold atoms in the single $\lvert\downarrow\rangle$ state, we can adjust the scattering length $a_{\downarrow\downarrow}$ to negative values, providing the attractive interaction required for the formation of Townes solitons.

%%%%%%%%%%%%%***Fig. 2***%%%%%%%%%%%%%%%%%%
 \begin{figure}[tbp]
\centerline{\includegraphics[scale=1.0]{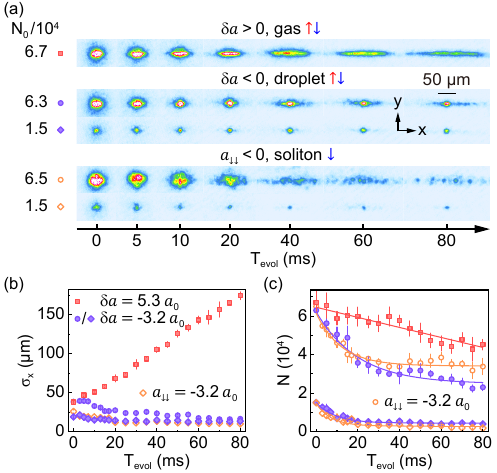}}
\caption{Experimental observation of 2D quantum droplets and Townes solitons. (a) Representative absorption images of spin-mixture gaseous BECs (solid square), quantum droplets (solid circle and solid diamond), and Townes solitons (open circle and open diamond) at different evolution times and atom numbers. The spin-mixture gaseous BECs and quantum droplets are prepared at effective mean-field scattering lengths of $\delta a=5.3\,a_0$ and $\delta a=-3.2\,a_0$, respectively, whereas the Townes solitons are realized in the single-spin condensates with a scattering length of $a_{\downarrow\downarrow}=-3.2\,a_0$. (b) and (c) show the time evolution of the cloud size along the $x$-direction and the atom number, respectively, for different states shown in (a). The solid lines in panel (c) are exponential fits to the atom-number decay.}
\label{fig2}
\end{figure}

%%%%%%%%%%%%%****Fig. 2***%%%%%%%%%%%%%%%%%%

%%%%%%%%%%%%%***Fig. 3***%%%%%%%%%%%%%%%%%%
 \begin{figure}[htbp]
\centerline{\includegraphics[scale=1.0]{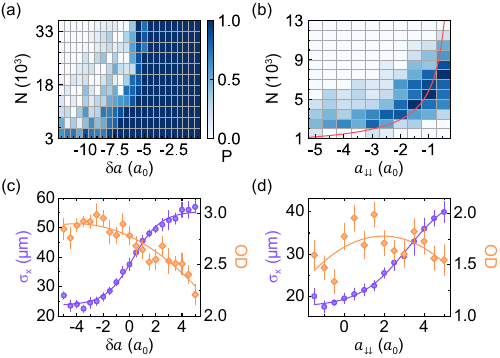}}
\caption{(a), (b) Phase diagrams for a single stable 2D quantum droplet and Townes soliton, respectively, as a function of the effective mean-field scattering length and atom number. The color of each grid cell indicates the probability of successfully preparing a single quantum droplet or Townes soliton, as determined from 10 experimental measurements. The solid red line in (b) represents the theoretical critical atom number of Townes solitons~\cite{SM}. (c), (d) Atomic cloud size and peak OD as functions of the effective mean-field scattering length for the two-spin mixture and the single-spin BECs, respectively. Both the measurements in (c) and (d) are performed after 20 ms of evolution, and the corresponding atom numbers are $N=2.2\times10^4$ and $N=9\times10^3$ in (c) and (d), respectively. The blue solid lines in (c) and (d) are fits to the cloud size $\sigma_x$, while the orange solid lines are fits to the OD. These lines are included as guides to the eye.}
\label{fig3}
\end{figure}
%%%%%%%%%%%%%****Fig. 3***%%%%%%%%%%%%%%%%%%

Our observations of quantum droplets and Townes solitons are presented in Fig.~\hyperref[fig2]{2}. Representative absorption-imaging pictures of ultracold atoms are presented in Fig.~\hyperref[fig2]{2(a)}. In the gas region with $\delta a>0$, the spin mixture expands anisotropically in the x-y plane owing to the anisotropic potential introduced by the accordion optical lattice~\cite{SM}. In sharp contrast, within the droplet region with $\delta a<0$, the spin mixture exhibits the pronounced self-binding property of quantum droplets. Specifically, the droplets maintain a compact size and a relatively high atomic number density over the entire observation period. We note that the 2D droplets prepared in our experiment exhibit much larger sizes and longer lifetimes than those previously reported quantum droplets~\cite{PRL.120.235301, Science359.301, PRL.120.135301}, as shown in Figs.~\hyperref[fig2]{2(b)} and \hyperref[fig2]{2(c)}. The 2D droplets observed in our experiments have a size of approximately $25\,\mu\mathrm{m}$ and a maximum lifetime exceeding $250\,\mathrm{ms}$~\cite{SM}.

For comparison, with the same quasi-2D geometry, we have also produced Townes solitons in the single $\lvert\downarrow\rangle$ state, which exhibit behaviors distinct from those of droplets. For a given negative scattering length, a gas with a large initial atom number undergoes fragmentation into multiple Townes solitons, as shown in Fig.~\hyperref[fig2]{2(a)}. By appropriately choosing the initial atom number, a single Townes soliton can instead be generated and maintained. The observation of Townes solitons confirms the 2D character of the system, thereby supporting the identification of the droplets formed in the same quasi-2D geometry as 2D quantum droplets.

The stability of droplets and solitons strongly depends on the scattering length and atom number. We have examined the parameter regions where a single self-bound droplet or soliton can stably exist, and the experimental results are presented in Figs.~\hyperref[fig3]{3(a)} and \hyperref[fig3]{3(b)}. We extensively vary the scattering length and atom number. After allowing the ultracold atoms to evolve in the quasi-2D geometry for 20 ms (droplets) or 30 ms (solitons), we check for the presence of single self-bound atomic clouds. In each small parameter region in Figs.~\hyperref[fig3]{3(a)} and \hyperref[fig3]{3(b)}, we repeat the experiment several times and determine the single-droplet or single-soliton formation probability. As shown in Fig.~\hyperref[fig3]{3(b)}, the solitons exhibit the characteristic behavior of Townes solitons, with maximum stability occurring near a critical atom number that is inversely proportional to the scattering length~\cite{PRL.125.250401, PRL.127.023603, SM}. When the atom number deviates from its critical value, the solitons become unstable, in agreement with the theoretical prediction~\cite{EPJST.225.2507}. In the two-spin mixture, stable quantum droplets can form over a much broader parameter region, without being restricted to the critical condition required for Townes solitons. In contrast to Townes solitons, the robust and stable formation of 2D quantum droplets can be attributed to the LHY effect, which takes an effectively 2D form under tight axial confinement~\cite{PRL.117.100401, PRA.98.051603, SM}.

We note that in 2D geometry, both the mean-field interaction and dispersion of the matter-wave packet are inversely proportional to the square of the wave-packet size. Consequently, the energy functional of the matter-wave packet can exhibit a flat profile as a function of the packet size, which supports the existence of metastable Townes solitons and endows them with scale invariance~\cite{SM, PRL.127.023603}. Introducing the LHY effect, which does not have a simple inverse-square relationship with size, will destroy the distinctive scale invariance of Townes solitons. Thus, Townes solitons cannot form in the spin mixture with LHY interaction.

%%%%%%%%%%%%%***Fig. 4***%%%%%%%%%%%%%%%%%%
 \begin{figure}[tbp]
\centerline{\includegraphics[scale=1.0]{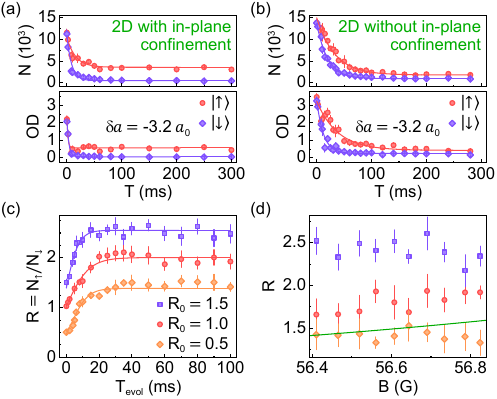}}
\caption{(a), (b) Lifetime of the spin mixture in a quasi-2D geometry with and without in-plane confinement. The effective mean-field scattering length is tuned to $\delta a=-3.2\,a_0$. While atoms in the mixture with in-plane confinement undergo rapid loss, particularly those in the $\lvert\downarrow\rangle$ state, the loss rate of the mixture without in-plane confinement is significantly reduced, leaving enough atoms to sustain a high-density quantum droplet. The solid lines in (a) and (b) are exponential fits to the experimental data. (c) The relative spin populations of the stable quantum droplets after thermal relaxation for different initial population ratios $R_0$. The solid lines are fits to the experimental data. (d) Relative spin populations of the quantum droplets as a function of Feshbach magnetic field. The measurements are performed after 60 ms of evolution in the quasi-2D geometry. The green solid line indicates the theoretically optimal population ratio.}
\label{fig4}
\end{figure}
%%%%%%%%%%%%%****Fig. 4***%%%%%%%%%%%%%%%%%%

Beyond the parameter regions shown in Figs.~\hyperref[fig3]{3(a)} and \hyperref[fig3]{3(b)}, we also examine whether droplets and solitons could exist in the repulsive mean-field regime. We adjust the effective scattering length $\delta a$ of the two-spin mixture and the scattering length $a_{\downarrow\downarrow}$ of atoms in the single $\lvert\downarrow\rangle$ state between positive and negative values, and then measure the size and peak optical depth (OD) of the atomic cloud. The measurement results are shown in Figs.~\hyperref[fig3]{3(c)} and \hyperref[fig3]{3(d)}. From the figure, we can see that droplets and solitons exist only in the attractive mean-field region with $\delta a<0$ and $a_{\downarrow\downarrow}<0$, respectively. In the repulsive regime, the atomic clouds remain gaseous BECs that expand in the quasi-2D geometry.

We further investigate the lifetime of droplets in the quasi-2D geometry. As a reference, we first measure the lifetimes of single-spin BECs in the quasi-2D geometry with in-plane confinement. The measured lifetimes are approximately $3.8\,\mathrm{s}$ for the $\lvert\uparrow\rangle$ state with $a_{\uparrow\uparrow}=33.7\,a_0$, and $0.34\,\mathrm{s}$ for the $\lvert\downarrow\rangle$ state with $a_{\downarrow\downarrow}=74.6\,a_0$~\cite{SM}. The substantially shorter lifetime of the $\lvert\downarrow\rangle$ condensates can be attributed to its larger three-body loss coefficient~\cite{PRL.120.235301, PRL.120.135301}. We then prepare a spin mixture in the droplet regime, with and without in-plane confinement, and monitor the evolution of the atom number and OD of both spin components, as shown in Figs.~\hyperref[fig4]{4(a)} and \hyperref[fig4]{4(b)}. In the presence of in-plane confinement, the rapid loss of the $\lvert\downarrow\rangle$ atoms limits the lifetime of the spin mixture to about $8\,\mathrm{ms}$. By contrast, with the in-plane confinement removed, the self-bound droplets can be spatially distinguished from the unbound background atoms, allowing us to count only the atoms remaining in the droplets. It is found that the atom-loss rate in the droplets is strongly suppressed owing to the reduced density of the unbound background atoms without confinement. Although the $\lvert\downarrow\rangle$ component still undergoes a rapid initial decay, a sufficient number of atoms remains in this component. Consequently, the quantum droplets remain long-lived, with lifetimes exceeding $250\,\mathrm{ms}$.

Previous experimental observations have shown that the population ratio $R = N_\uparrow /N_\downarrow$ of stable quantum droplets in binary Bose mixtures approaches the optimal value $\sqrt{a_{\downarrow\downarrow}/a_{\uparrow\uparrow}}$~\cite{PRL.120.235301, PRL.120.135301, PRR.3.033247}. Here, we demonstrate that the equilibrium population ratio can deviate substantially from this value. As shown in Fig.~\hyperref[fig4]{4(c)}, we prepare initial population ratios of $R_0=0.5, 1.0, \mathrm{and}~1.5$ by varying the RF pulse duration and monitor their subsequent evolution in the quasi-2D geometry. At $B = 56.59\,\mathrm{G}$ with scattering lengths of $(a_{\uparrow\uparrow}, a_{\downarrow\downarrow}, a_{\uparrow\downarrow})/a_0=(33.7, 74.6, -53.4)$, these ratios rapidly relax to steady-state values of about $1.5$, $2$, and $2.5$, respectively, with the latter two deviating substantially from the optimal value of $1.5$. In Fig.~\hyperref[fig4]{4(d)}, by tuning the magnetic field to modify the scattering lengths, we further demonstrate that such deviations persist over a broad range of interaction parameters, suggesting that they are a generic feature of stable droplets. These observations agree with recent theoretical predictions that the optimal population ratio is not a prerequisite for droplet formation~\cite{PRR.5.033167, PRR.6.013209, PRA.110.023324, PRA.111.023309}, and provide direct experimental support for this emerging picture.

In summary, we have realized both quantum droplets in a binary Bose mixture and Townes solitons in single-component BECs within the same quasi-2D geometry with tight axial confinement. The quantum droplets exhibit remarkable stability, with lifetimes nearly an order of magnitude longer than those observed in 3D free space. The extended lifetime in the quasi-2D geometry enables the realization of self-bound Moir\'{e} droplets in a twisted-bilayer optical lattice~\cite{arXiv:2609.28913}, where the interspecies attraction provides an alternative to the Rabi field in~\cite{Nature615.231} for coupling the two layers. Since the LHY correction, atomic interactions, and matter-wave dispersion are all strongly influenced by dimensionality, our results provide new opportunities for exploring the properties and dynamics of quantum liquids in low-dimensional systems.

\begin{acknowledgments}

This research is supported by the National Key Research and Development Program of China (Grants No. 2021YFA1401700, 2022YFA1404101), Quantum Science and Technology-National Science and Technology Major Project (Grant No. 2021ZD0302003), the National Natural Science Foundation of China (Grants No. U23A6004, 12488301, 92476001, 12374245, 12474266, 12474252, 12322409, 12004229, 92576205, 12504336), and the Fundamental Research Program of Shanxi Province (Grant No. 202403021221001).

\end{acknowledgments}


\begin{thebibliography}{61}%
\makeatletter
\providecommand \@ifxundefined [1]{%
 \@ifx{#1\undefined}
}%
\providecommand \@ifnum [1]{%
 \ifnum #1\expandafter \@firstoftwo
 \else \expandafter \@secondoftwo
 \fi
}%
\providecommand \@ifx [1]{%
 \ifx #1\expandafter \@firstoftwo
 \else \expandafter \@secondoftwo
 \fi
}%
\providecommand \natexlab [1]{#1}%
\providecommand \enquote  [1]{``#1''}%
\providecommand \bibnamefont  [1]{#1}%
\providecommand \bibfnamefont [1]{#1}%
\providecommand \citenamefont [1]{#1}%
\providecommand \href@noop [0]{\@secondoftwo}%
\providecommand \href [0]{\begingroup \@sanitize@url \@href}%
\providecommand \@href[1]{\@@startlink{#1}\@@href}%
\providecommand \@@href[1]{\endgroup#1\@@endlink}%
\providecommand \@sanitize@url [0]{\catcode `\\12\catcode `\$12\catcode
  `\&12\catcode `\#12\catcode `\^12\catcode `\_12\catcode `\%12\relax}%
\providecommand \@@startlink[1]{}%
\providecommand \@@endlink[0]{}%
\providecommand \url  [0]{\begingroup\@sanitize@url \@url }%
\providecommand \@url [1]{\endgroup\@href {#1}{\urlprefix }}%
\providecommand \urlprefix  [0]{URL }%
\providecommand \Eprint [0]{\href }%
\providecommand \doibase [0]{https://doi.org/}%
\providecommand \selectlanguage [0]{\@gobble}%
\providecommand \bibinfo  [0]{\@secondoftwo}%
\providecommand \bibfield  [0]{\@secondoftwo}%
\providecommand \translation [1]{[#1]}%
\providecommand \BibitemOpen [0]{}%
\providecommand \bibitemStop [0]{}%
\providecommand \bibitemNoStop [0]{.\EOS\space}%
\providecommand \EOS [0]{\spacefactor3000\relax}%
\providecommand \BibitemShut  [1]{\csname bibitem#1\endcsname}%
\let\auto@bib@innerbib\@empty
%</preamble>
\bibitem [{\citenamefont {Petrov}(2015)}]{PRL.115.155302}%
  \BibitemOpen
  \bibfield  {author} {\bibinfo {author} {\bibfnamefont {D.~S.}\ \bibnamefont
  {Petrov}},\ }\bibfield  {title} {\bibinfo {title} {Quantum mechanical
  stabilization of a collapsing {Bose-Bose} mixture},\ }\href
  {https://doi.org/https://doi.org/10.1103/PhysRevLett.115.155302} {\bibfield
  {journal} {\bibinfo  {journal} {Phys. Rev. Lett.}\ }\textbf {\bibinfo
  {volume} {115}},\ \bibinfo {pages} {155302} (\bibinfo {year}
  {2015})}\BibitemShut {NoStop}%
\bibitem [{\citenamefont {Cabrera}\ \emph {et~al.}(2018)\citenamefont
  {Cabrera}, \citenamefont {Tanzi}, \citenamefont {Sanz}, \citenamefont
  {Naylor}, \citenamefont {Thomas}, \citenamefont {Cheiney},\ and\
  \citenamefont {Tarruell}}]{Science359.301}%
  \BibitemOpen
  \bibfield  {author} {\bibinfo {author} {\bibfnamefont {C.~R.}\ \bibnamefont
  {Cabrera}}, \bibinfo {author} {\bibfnamefont {L.}~\bibnamefont {Tanzi}},
  \bibinfo {author} {\bibfnamefont {J.}~\bibnamefont {Sanz}}, \bibinfo {author}
  {\bibfnamefont {B.}~\bibnamefont {Naylor}}, \bibinfo {author} {\bibfnamefont
  {P.}~\bibnamefont {Thomas}}, \bibinfo {author} {\bibfnamefont
  {P.}~\bibnamefont {Cheiney}},\ and\ \bibinfo {author} {\bibfnamefont
  {L.}~\bibnamefont {Tarruell}},\ }\bibfield  {title} {\bibinfo {title}
  {Quantum liquid droplets in a mixture of {Bose-Einstein} condensates},\
  }\href {https://doi.org/10.1126/science.aao5686} {\bibfield  {journal}
  {\bibinfo  {journal} {Science}\ }\textbf {\bibinfo {volume} {359}},\ \bibinfo
  {pages} {301} (\bibinfo {year} {2018})}\BibitemShut {NoStop}%
\bibitem [{\citenamefont {Semeghini}\ \emph {et~al.}(2018)\citenamefont
  {Semeghini}, \citenamefont {Ferioli}, \citenamefont {Masi}, \citenamefont
  {Mazzinghi}, \citenamefont {Wolswijk}, \citenamefont {Minardi}, \citenamefont
  {Modugno}, \citenamefont {Modugno}, \citenamefont {Inguscio},\ and\
  \citenamefont {Fattori}}]{PRL.120.235301}%
  \BibitemOpen
  \bibfield  {author} {\bibinfo {author} {\bibfnamefont {G.}~\bibnamefont
  {Semeghini}}, \bibinfo {author} {\bibfnamefont {G.}~\bibnamefont {Ferioli}},
  \bibinfo {author} {\bibfnamefont {L.}~\bibnamefont {Masi}}, \bibinfo {author}
  {\bibfnamefont {C.}~\bibnamefont {Mazzinghi}}, \bibinfo {author}
  {\bibfnamefont {L.}~\bibnamefont {Wolswijk}}, \bibinfo {author}
  {\bibfnamefont {F.}~\bibnamefont {Minardi}}, \bibinfo {author} {\bibfnamefont
  {M.}~\bibnamefont {Modugno}}, \bibinfo {author} {\bibfnamefont
  {G.}~\bibnamefont {Modugno}}, \bibinfo {author} {\bibfnamefont
  {M.}~\bibnamefont {Inguscio}},\ and\ \bibinfo {author} {\bibfnamefont
  {M.}~\bibnamefont {Fattori}},\ }\bibfield  {title} {\bibinfo {title}
  {Self-bound quantum droplets of atomic mixtures in free space},\ }\href
  {https://doi.org/https://doi.org/10.1103/PhysRevLett.120.235301} {\bibfield
  {journal} {\bibinfo  {journal} {Phys. Rev. Lett.}\ }\textbf {\bibinfo
  {volume} {120}},\ \bibinfo {pages} {235301} (\bibinfo {year}
  {2018})}\BibitemShut {NoStop}%
\bibitem [{\citenamefont {Cheiney}\ \emph {et~al.}(2018)\citenamefont
  {Cheiney}, \citenamefont {Cabrera}, \citenamefont {Sanz}, \citenamefont
  {Naylor}, \citenamefont {Tanzi},\ and\ \citenamefont
  {Tarruell}}]{PRL.120.135301}%
  \BibitemOpen
  \bibfield  {author} {\bibinfo {author} {\bibfnamefont {P.}~\bibnamefont
  {Cheiney}}, \bibinfo {author} {\bibfnamefont {C.~R.}\ \bibnamefont
  {Cabrera}}, \bibinfo {author} {\bibfnamefont {J.}~\bibnamefont {Sanz}},
  \bibinfo {author} {\bibfnamefont {B.}~\bibnamefont {Naylor}}, \bibinfo
  {author} {\bibfnamefont {L.}~\bibnamefont {Tanzi}},\ and\ \bibinfo {author}
  {\bibfnamefont {L.}~\bibnamefont {Tarruell}},\ }\bibfield  {title} {\bibinfo
  {title} {Bright soliton to quantum droplet transition in a mixture of
  {Bose-Einstein} condensates},\ }\href
  {https://doi.org/https://doi.org/10.1103/PhysRevLett.120.135301} {\bibfield
  {journal} {\bibinfo  {journal} {Phys. Rev. Lett.}\ }\textbf {\bibinfo
  {volume} {120}},\ \bibinfo {pages} {135301} (\bibinfo {year}
  {2018})}\BibitemShut {NoStop}%
\bibitem [{\citenamefont {D'Errico}\ \emph {et~al.}(2019)\citenamefont
  {D'Errico}, \citenamefont {Burchianti}, \citenamefont {Prevedelli},
  \citenamefont {Salasnich}, \citenamefont {Ancilotto}, \citenamefont
  {Modugno}, \citenamefont {Minardi},\ and\ \citenamefont
  {Fort}}]{PRR.1.033155}%
  \BibitemOpen
  \bibfield  {author} {\bibinfo {author} {\bibfnamefont {C.}~\bibnamefont
  {D'Errico}}, \bibinfo {author} {\bibfnamefont {A.}~\bibnamefont
  {Burchianti}}, \bibinfo {author} {\bibfnamefont {M.}~\bibnamefont
  {Prevedelli}}, \bibinfo {author} {\bibfnamefont {L.}~\bibnamefont
  {Salasnich}}, \bibinfo {author} {\bibfnamefont {F.}~\bibnamefont
  {Ancilotto}}, \bibinfo {author} {\bibfnamefont {M.}~\bibnamefont {Modugno}},
  \bibinfo {author} {\bibfnamefont {F.}~\bibnamefont {Minardi}},\ and\ \bibinfo
  {author} {\bibfnamefont {C.}~\bibnamefont {Fort}},\ }\bibfield  {title}
  {\bibinfo {title} {Observation of quantum droplets in a heteronuclear bosonic
  mixture},\ }\href {https://doi.org/10.1103/PhysRevResearch.1.033155}
  {\bibfield  {journal} {\bibinfo  {journal} {Phys. Rev. Res.}\ }\textbf
  {\bibinfo {volume} {1}},\ \bibinfo {pages} {033155} (\bibinfo {year}
  {2019})}\BibitemShut {NoStop}%
\bibitem [{\citenamefont {Guo}\ \emph {et~al.}(2021)\citenamefont {Guo},
  \citenamefont {Jia}, \citenamefont {Li}, \citenamefont {Ma}, \citenamefont
  {Hutson}, \citenamefont {Cui},\ and\ \citenamefont {Wang}}]{PRR.3.033247}%
  \BibitemOpen
  \bibfield  {author} {\bibinfo {author} {\bibfnamefont {Z.}~\bibnamefont
  {Guo}}, \bibinfo {author} {\bibfnamefont {F.}~\bibnamefont {Jia}}, \bibinfo
  {author} {\bibfnamefont {L.}~\bibnamefont {Li}}, \bibinfo {author}
  {\bibfnamefont {Y.}~\bibnamefont {Ma}}, \bibinfo {author} {\bibfnamefont
  {J.~M.}\ \bibnamefont {Hutson}}, \bibinfo {author} {\bibfnamefont
  {X.}~\bibnamefont {Cui}},\ and\ \bibinfo {author} {\bibfnamefont
  {D.}~\bibnamefont {Wang}},\ }\bibfield  {title} {\bibinfo {title}
  {{Lee-Huang-Yang} effects in the ultracold mixture of $^{23}\mathrm{Na}$ and
  $^{87}\mathrm{Rb}$ with attractive interspecies interactions},\ }\href
  {https://doi.org/10.1103/PhysRevResearch.3.033247} {\bibfield  {journal}
  {\bibinfo  {journal} {Phys. Rev. Res.}\ }\textbf {\bibinfo {volume} {3}},\
  \bibinfo {pages} {033247} (\bibinfo {year} {2021})}\BibitemShut {NoStop}%
\bibitem [{\citenamefont {B{\"o}ttcher}\ \emph {et~al.}(2021)\citenamefont
  {B{\"o}ttcher}, \citenamefont {Schmidt}, \citenamefont {Hertkorn},
  \citenamefont {Ng}, \citenamefont {Graham}, \citenamefont {Guo},
  \citenamefont {Langen},\ and\ \citenamefont {Pfau}}]{RPP.84.012403}%
  \BibitemOpen
  \bibfield  {author} {\bibinfo {author} {\bibfnamefont {F.}~\bibnamefont
  {B{\"o}ttcher}}, \bibinfo {author} {\bibfnamefont {J.-N.}\ \bibnamefont
  {Schmidt}}, \bibinfo {author} {\bibfnamefont {J.}~\bibnamefont {Hertkorn}},
  \bibinfo {author} {\bibfnamefont {K.~S.~H.}\ \bibnamefont {Ng}}, \bibinfo
  {author} {\bibfnamefont {S.~D.}\ \bibnamefont {Graham}}, \bibinfo {author}
  {\bibfnamefont {M.}~\bibnamefont {Guo}}, \bibinfo {author} {\bibfnamefont
  {T.}~\bibnamefont {Langen}},\ and\ \bibinfo {author} {\bibfnamefont
  {T.}~\bibnamefont {Pfau}},\ }\bibfield  {title} {\bibinfo {title} {New states
  of matter with fine-tuned interactions: Quantum droplets and dipolar
  supersolids},\ }\href {https://doi.org/10.1088/1361-6633/abc9ab} {\bibfield
  {journal} {\bibinfo  {journal} {Rep. Prog. Phys.}\ }\textbf {\bibinfo
  {volume} {84}},\ \bibinfo {pages} {012403} (\bibinfo {year}
  {2021})}\BibitemShut {NoStop}%
\bibitem [{\citenamefont {Tanzi}\ \emph {et~al.}(2019)\citenamefont {Tanzi},
  \citenamefont {Roccuzzo}, \citenamefont {Lucioni}, \citenamefont {Fam{\`a}},
  \citenamefont {Fioretti}, \citenamefont {Gabbanini}, \citenamefont {Modugno},
  \citenamefont {Recati},\ and\ \citenamefont {Stringari}}]{Nature574.382}%
  \BibitemOpen
  \bibfield  {author} {\bibinfo {author} {\bibfnamefont {L.}~\bibnamefont
  {Tanzi}}, \bibinfo {author} {\bibfnamefont {S.~M.}\ \bibnamefont {Roccuzzo}},
  \bibinfo {author} {\bibfnamefont {E.}~\bibnamefont {Lucioni}}, \bibinfo
  {author} {\bibfnamefont {F.}~\bibnamefont {Fam{\`a}}}, \bibinfo {author}
  {\bibfnamefont {A.}~\bibnamefont {Fioretti}}, \bibinfo {author}
  {\bibfnamefont {C.}~\bibnamefont {Gabbanini}}, \bibinfo {author}
  {\bibfnamefont {G.}~\bibnamefont {Modugno}}, \bibinfo {author} {\bibfnamefont
  {A.}~\bibnamefont {Recati}},\ and\ \bibinfo {author} {\bibfnamefont
  {S.}~\bibnamefont {Stringari}},\ }\bibfield  {title} {\bibinfo {title}
  {Supersolid symmetry breaking from compressional oscillations in a dipolar
  quantum gas},\ }\href
  {https://doi.org/https://doi.org/10.1038/s41586-019-1568-6} {\bibfield
  {journal} {\bibinfo  {journal} {Nature}\ }\textbf {\bibinfo {volume} {574}},\
  \bibinfo {pages} {382} (\bibinfo {year} {2019})}\BibitemShut {NoStop}%
\bibitem [{\citenamefont {Pan}\ \emph {et~al.}(2022)\citenamefont {Pan},
  \citenamefont {Yi},\ and\ \citenamefont {Shi}}]{PRR.4.043018}%
  \BibitemOpen
  \bibfield  {author} {\bibinfo {author} {\bibfnamefont {J.}~\bibnamefont
  {Pan}}, \bibinfo {author} {\bibfnamefont {S.}~\bibnamefont {Yi}},\ and\
  \bibinfo {author} {\bibfnamefont {T.}~\bibnamefont {Shi}},\ }\bibfield
  {title} {\bibinfo {title} {Quantum phases of self-bound droplets of
  {Bose-Bose} mixtures},\ }\href
  {https://doi.org/10.1103/PhysRevResearch.4.043018} {\bibfield  {journal}
  {\bibinfo  {journal} {Phys. Rev. Res.}\ }\textbf {\bibinfo {volume} {4}},\
  \bibinfo {pages} {043018} (\bibinfo {year} {2022})}\BibitemShut {NoStop}%
\bibitem [{\citenamefont {Sachdeva}\ \emph {et~al.}(2020)\citenamefont
  {Sachdeva}, \citenamefont {Tengstrand},\ and\ \citenamefont
  {Reimann}}]{PRA.102.043304}%
  \BibitemOpen
  \bibfield  {author} {\bibinfo {author} {\bibfnamefont {R.}~\bibnamefont
  {Sachdeva}}, \bibinfo {author} {\bibfnamefont {M.~N.}\ \bibnamefont
  {Tengstrand}},\ and\ \bibinfo {author} {\bibfnamefont {S.~M.}\ \bibnamefont
  {Reimann}},\ }\bibfield  {title} {\bibinfo {title} {Self-bound supersolid
  stripe phase in binary {Bose-Einstein} condensates},\ }\href
  {https://doi.org/10.1103/PhysRevA.102.043304} {\bibfield  {journal} {\bibinfo
   {journal} {Phys. Rev. A}\ }\textbf {\bibinfo {volume} {102}},\ \bibinfo
  {pages} {043304} (\bibinfo {year} {2020})}\BibitemShut {NoStop}%
\bibitem [{\citenamefont {S\'anchez-Baena}\ \emph {et~al.}(2020)\citenamefont
  {S\'anchez-Baena}, \citenamefont {Boronat},\ and\ \citenamefont
  {Mazzanti}}]{PRA.102.053308}%
  \BibitemOpen
  \bibfield  {author} {\bibinfo {author} {\bibfnamefont {J.}~\bibnamefont
  {S\'anchez-Baena}}, \bibinfo {author} {\bibfnamefont {J.}~\bibnamefont
  {Boronat}},\ and\ \bibinfo {author} {\bibfnamefont {F.}~\bibnamefont
  {Mazzanti}},\ }\bibfield  {title} {\bibinfo {title} {Supersolid striped
  droplets in a {Raman} spin-orbit-coupled system},\ }\href
  {https://doi.org/10.1103/PhysRevA.102.053308} {\bibfield  {journal} {\bibinfo
   {journal} {Phys. Rev. A}\ }\textbf {\bibinfo {volume} {102}},\ \bibinfo
  {pages} {053308} (\bibinfo {year} {2020})}\BibitemShut {NoStop}%
\bibitem [{\citenamefont {Ferioli}\ \emph {et~al.}(2019)\citenamefont
  {Ferioli}, \citenamefont {Semeghini}, \citenamefont {Masi}, \citenamefont
  {Giusti}, \citenamefont {Modugno}, \citenamefont {Inguscio}, \citenamefont
  {Gallem\'{\i}}, \citenamefont {Recati},\ and\ \citenamefont
  {Fattori}}]{PRL.122.090401}%
  \BibitemOpen
  \bibfield  {author} {\bibinfo {author} {\bibfnamefont {G.}~\bibnamefont
  {Ferioli}}, \bibinfo {author} {\bibfnamefont {G.}~\bibnamefont {Semeghini}},
  \bibinfo {author} {\bibfnamefont {L.}~\bibnamefont {Masi}}, \bibinfo {author}
  {\bibfnamefont {G.}~\bibnamefont {Giusti}}, \bibinfo {author} {\bibfnamefont
  {G.}~\bibnamefont {Modugno}}, \bibinfo {author} {\bibfnamefont
  {M.}~\bibnamefont {Inguscio}}, \bibinfo {author} {\bibfnamefont
  {A.}~\bibnamefont {Gallem\'{\i}}}, \bibinfo {author} {\bibfnamefont
  {A.}~\bibnamefont {Recati}},\ and\ \bibinfo {author} {\bibfnamefont
  {M.}~\bibnamefont {Fattori}},\ }\bibfield  {title} {\bibinfo {title}
  {Collisions of self-bound quantum droplets},\ }\href
  {https://doi.org/10.1103/PhysRevLett.122.090401} {\bibfield  {journal}
  {\bibinfo  {journal} {Phys. Rev. Lett.}\ }\textbf {\bibinfo {volume} {122}},\
  \bibinfo {pages} {090401} (\bibinfo {year} {2019})}\BibitemShut {NoStop}%
\bibitem [{\citenamefont {Dong}\ and\ \citenamefont
  {Kartashov}(2021)}]{PRL.126.244101}%
  \BibitemOpen
  \bibfield  {author} {\bibinfo {author} {\bibfnamefont {L.}~\bibnamefont
  {Dong}}\ and\ \bibinfo {author} {\bibfnamefont {Y.~V.}\ \bibnamefont
  {Kartashov}},\ }\bibfield  {title} {\bibinfo {title} {Rotating
  multidimensional quantum droplets},\ }\href
  {https://doi.org/10.1103/PhysRevLett.126.244101} {\bibfield  {journal}
  {\bibinfo  {journal} {Phys. Rev. Lett.}\ }\textbf {\bibinfo {volume} {126}},\
  \bibinfo {pages} {244101} (\bibinfo {year} {2021})}\BibitemShut {NoStop}%
\bibitem [{\citenamefont {Tengstrand}\ \emph {et~al.}(2019)\citenamefont
  {Tengstrand}, \citenamefont {St\"urmer}, \citenamefont {Karabulut},\ and\
  \citenamefont {Reimann}}]{PRL.123.160405}%
  \BibitemOpen
  \bibfield  {author} {\bibinfo {author} {\bibfnamefont {M.~N.}\ \bibnamefont
  {Tengstrand}}, \bibinfo {author} {\bibfnamefont {P.}~\bibnamefont
  {St\"urmer}}, \bibinfo {author} {\bibfnamefont {E.~{\"O}.}\ \bibnamefont
  {Karabulut}},\ and\ \bibinfo {author} {\bibfnamefont {S.~M.}\ \bibnamefont
  {Reimann}},\ }\bibfield  {title} {\bibinfo {title} {Rotating binary
  {Bose-Einstein} condensates and vortex clusters in quantum droplets},\ }\href
  {https://doi.org/10.1103/PhysRevLett.123.160405} {\bibfield  {journal}
  {\bibinfo  {journal} {Phys. Rev. Lett.}\ }\textbf {\bibinfo {volume} {123}},\
  \bibinfo {pages} {160405} (\bibinfo {year} {2019})}\BibitemShut {NoStop}%
\bibitem [{\citenamefont {Petrov}\ and\ \citenamefont
  {Astrakharchik}(2016)}]{PRL.117.100401}%
  \BibitemOpen
  \bibfield  {author} {\bibinfo {author} {\bibfnamefont {D.~S.}\ \bibnamefont
  {Petrov}}\ and\ \bibinfo {author} {\bibfnamefont {G.~E.}\ \bibnamefont
  {Astrakharchik}},\ }\bibfield  {title} {\bibinfo {title} {Ultradilute
  low-dimensional liquids},\ }\href
  {https://doi.org/10.1103/PhysRevLett.117.100401} {\bibfield  {journal}
  {\bibinfo  {journal} {Phys. Rev. Lett.}\ }\textbf {\bibinfo {volume} {117}},\
  \bibinfo {pages} {100401} (\bibinfo {year} {2016})}\BibitemShut {NoStop}%
\bibitem [{\citenamefont {Ilg}\ \emph {et~al.}(2018)\citenamefont {Ilg},
  \citenamefont {Kumlin}, \citenamefont {Santos}, \citenamefont {Petrov},\ and\
  \citenamefont {B\"uchler}}]{PRA.98.051604}%
  \BibitemOpen
  \bibfield  {author} {\bibinfo {author} {\bibfnamefont {T.}~\bibnamefont
  {Ilg}}, \bibinfo {author} {\bibfnamefont {J.}~\bibnamefont {Kumlin}},
  \bibinfo {author} {\bibfnamefont {L.}~\bibnamefont {Santos}}, \bibinfo
  {author} {\bibfnamefont {D.~S.}\ \bibnamefont {Petrov}},\ and\ \bibinfo
  {author} {\bibfnamefont {H.~P.}\ \bibnamefont {B\"uchler}},\ }\bibfield
  {title} {\bibinfo {title} {Dimensional crossover for the beyond-mean-field
  correction in {Bose} gases},\ }\href
  {https://doi.org/10.1103/PhysRevA.98.051604} {\bibfield  {journal} {\bibinfo
  {journal} {Phys. Rev. A}\ }\textbf {\bibinfo {volume} {98}},\ \bibinfo
  {pages} {051604(R)} (\bibinfo {year} {2018})}\BibitemShut {NoStop}%
\bibitem [{\citenamefont {Luo}\ \emph {et~al.}(2020)\citenamefont {Luo},
  \citenamefont {Pang}, \citenamefont {Liu}, \citenamefont {Li},\ and\
  \citenamefont {Malomed}}]{FP.16.32201}%
  \BibitemOpen
  \bibfield  {author} {\bibinfo {author} {\bibfnamefont {Z.-H.}\ \bibnamefont
  {Luo}}, \bibinfo {author} {\bibfnamefont {W.}~\bibnamefont {Pang}}, \bibinfo
  {author} {\bibfnamefont {B.}~\bibnamefont {Liu}}, \bibinfo {author}
  {\bibfnamefont {Y.-Y.}\ \bibnamefont {Li}},\ and\ \bibinfo {author}
  {\bibfnamefont {B.~A.}\ \bibnamefont {Malomed}},\ }\bibfield  {title}
  {\bibinfo {title} {A new form of liquid matter: Quantum droplets},\ }\href
  {https://doi.org/10.1007/s11467-020-1020-2} {\bibfield  {journal} {\bibinfo
  {journal} {Front. Phys.}\ }\textbf {\bibinfo {volume} {16}},\ \bibinfo
  {pages} {32201} (\bibinfo {year} {2020})}\BibitemShut {NoStop}%
\bibitem [{\citenamefont {Parisi}\ \emph {et~al.}(2019)\citenamefont {Parisi},
  \citenamefont {Astrakharchik},\ and\ \citenamefont
  {Giorgini}}]{PRL.122.105302}%
  \BibitemOpen
  \bibfield  {author} {\bibinfo {author} {\bibfnamefont {L.}~\bibnamefont
  {Parisi}}, \bibinfo {author} {\bibfnamefont {G.~E.}\ \bibnamefont
  {Astrakharchik}},\ and\ \bibinfo {author} {\bibfnamefont {S.}~\bibnamefont
  {Giorgini}},\ }\bibfield  {title} {\bibinfo {title} {Liquid state of
  one-dimensional {Bose} mixtures: A quantum {Monte Carlo} study},\ }\href
  {https://doi.org/10.1103/PhysRevLett.122.105302} {\bibfield  {journal}
  {\bibinfo  {journal} {Phys. Rev. Lett.}\ }\textbf {\bibinfo {volume} {122}},\
  \bibinfo {pages} {105302} (\bibinfo {year} {2019})}\BibitemShut {NoStop}%
\bibitem [{\citenamefont {Spada}\ \emph {et~al.}(2024)\citenamefont {Spada},
  \citenamefont {Pilati},\ and\ \citenamefont {Giorgini}}]{PRL.133.083401}%
  \BibitemOpen
  \bibfield  {author} {\bibinfo {author} {\bibfnamefont {G.}~\bibnamefont
  {Spada}}, \bibinfo {author} {\bibfnamefont {S.}~\bibnamefont {Pilati}},\ and\
  \bibinfo {author} {\bibfnamefont {S.}~\bibnamefont {Giorgini}},\ }\bibfield
  {title} {\bibinfo {title} {Quantum droplets in two-dimensional {Bose}
  mixtures at finite temperature},\ }\href
  {https://doi.org/10.1103/PhysRevLett.133.083401} {\bibfield  {journal}
  {\bibinfo  {journal} {Phys. Rev. Lett.}\ }\textbf {\bibinfo {volume} {133}},\
  \bibinfo {pages} {083401} (\bibinfo {year} {2024})}\BibitemShut {NoStop}%
\bibitem [{\citenamefont {Cui}\ and\ \citenamefont {Ma}(2021)}]{PRR.3.012027}%
  \BibitemOpen
  \bibfield  {author} {\bibinfo {author} {\bibfnamefont {X.}~\bibnamefont
  {Cui}}\ and\ \bibinfo {author} {\bibfnamefont {Y.}~\bibnamefont {Ma}},\
  }\bibfield  {title} {\bibinfo {title} {Droplet under confinement: Competition
  and coexistence with a soliton bound state},\ }\href
  {https://doi.org/10.1103/PhysRevResearch.3.L012027} {\bibfield  {journal}
  {\bibinfo  {journal} {Phys. Rev. Res.}\ }\textbf {\bibinfo {volume} {3}},\
  \bibinfo {pages} {L012027} (\bibinfo {year} {2021})}\BibitemShut {NoStop}%
\bibitem [{\citenamefont {Zin}\ \emph {et~al.}(2018)\citenamefont {Zin},
  \citenamefont {Pylak}, \citenamefont {Wasak}, \citenamefont {Gajda},\ and\
  \citenamefont {Idziaszek}}]{PRA.98.051603}%
  \BibitemOpen
  \bibfield  {author} {\bibinfo {author} {\bibfnamefont {P.}~\bibnamefont
  {Zin}}, \bibinfo {author} {\bibfnamefont {M.}~\bibnamefont {Pylak}}, \bibinfo
  {author} {\bibfnamefont {T.}~\bibnamefont {Wasak}}, \bibinfo {author}
  {\bibfnamefont {M.}~\bibnamefont {Gajda}},\ and\ \bibinfo {author}
  {\bibfnamefont {Z.}~\bibnamefont {Idziaszek}},\ }\bibfield  {title} {\bibinfo
  {title} {Quantum {Bose-Bose} droplets at a dimensional crossover},\ }\href
  {https://doi.org/10.1103/PhysRevA.98.051603} {\bibfield  {journal} {\bibinfo
  {journal} {Phys. Rev. A}\ }\textbf {\bibinfo {volume} {98}},\ \bibinfo
  {pages} {051603(R)} (\bibinfo {year} {2018})}\BibitemShut {NoStop}%
\bibitem [{\citenamefont {Schmitt}\ \emph {et~al.}(2016)\citenamefont
  {Schmitt}, \citenamefont {Wenzel}, \citenamefont {B{\"o}ttcher},
  \citenamefont {Ferrier-Barbut},\ and\ \citenamefont {Pfau}}]{Nature539.259}%
  \BibitemOpen
  \bibfield  {author} {\bibinfo {author} {\bibfnamefont {M.}~\bibnamefont
  {Schmitt}}, \bibinfo {author} {\bibfnamefont {M.}~\bibnamefont {Wenzel}},
  \bibinfo {author} {\bibfnamefont {F.}~\bibnamefont {B{\"o}ttcher}}, \bibinfo
  {author} {\bibfnamefont {I.}~\bibnamefont {Ferrier-Barbut}},\ and\ \bibinfo
  {author} {\bibfnamefont {T.}~\bibnamefont {Pfau}},\ }\bibfield  {title}
  {\bibinfo {title} {Self-bound droplets of a dilute magnetic quantum liquid},\
  }\href {https://doi.org/10.1038/nature20126} {\bibfield  {journal} {\bibinfo
  {journal} {Nature}\ }\textbf {\bibinfo {volume} {539}},\ \bibinfo {pages}
  {259} (\bibinfo {year} {2016})}\BibitemShut {NoStop}%
\bibitem [{\citenamefont {Chomaz}\ \emph {et~al.}(2016)\citenamefont {Chomaz},
  \citenamefont {Baier}, \citenamefont {Petter}, \citenamefont {Mark},
  \citenamefont {W\"achtler}, \citenamefont {Santos},\ and\ \citenamefont
  {Ferlaino}}]{PRX.6.041039}%
  \BibitemOpen
  \bibfield  {author} {\bibinfo {author} {\bibfnamefont {L.}~\bibnamefont
  {Chomaz}}, \bibinfo {author} {\bibfnamefont {S.}~\bibnamefont {Baier}},
  \bibinfo {author} {\bibfnamefont {D.}~\bibnamefont {Petter}}, \bibinfo
  {author} {\bibfnamefont {M.~J.}\ \bibnamefont {Mark}}, \bibinfo {author}
  {\bibfnamefont {F.}~\bibnamefont {W\"achtler}}, \bibinfo {author}
  {\bibfnamefont {L.}~\bibnamefont {Santos}},\ and\ \bibinfo {author}
  {\bibfnamefont {F.}~\bibnamefont {Ferlaino}},\ }\bibfield  {title} {\bibinfo
  {title} {Quantum-fluctuation-driven crossover from a dilute {Bose-Einstein}
  condensate to a macrodroplet in a dipolar quantum fluid},\ }\href
  {https://doi.org/10.1103/PhysRevX.6.041039} {\bibfield  {journal} {\bibinfo
  {journal} {Phys. Rev. X}\ }\textbf {\bibinfo {volume} {6}},\ \bibinfo {pages}
  {041039} (\bibinfo {year} {2016})}\BibitemShut {NoStop}%
\bibitem [{\citenamefont {Skov}\ \emph {et~al.}(2021)\citenamefont {Skov},
  \citenamefont {Skou}, \citenamefont {J\o{}rgensen},\ and\ \citenamefont
  {Arlt}}]{PRL.126.230404}%
  \BibitemOpen
  \bibfield  {author} {\bibinfo {author} {\bibfnamefont {T.~G.}\ \bibnamefont
  {Skov}}, \bibinfo {author} {\bibfnamefont {M.~G.}\ \bibnamefont {Skou}},
  \bibinfo {author} {\bibfnamefont {N.~B.}\ \bibnamefont {J\o{}rgensen}},\ and\
  \bibinfo {author} {\bibfnamefont {J.~J.}\ \bibnamefont {Arlt}},\ }\bibfield
  {title} {\bibinfo {title} {Observation of a {Lee-Huang-Yang} fluid},\ }\href
  {https://doi.org/10.1103/PhysRevLett.126.230404} {\bibfield  {journal}
  {\bibinfo  {journal} {Phys. Rev. Lett.}\ }\textbf {\bibinfo {volume} {126}},\
  \bibinfo {pages} {230404} (\bibinfo {year} {2021})}\BibitemShut {NoStop}%
\bibitem [{\citenamefont {J\o{}rgensen}\ \emph {et~al.}(2018)\citenamefont
  {J\o{}rgensen}, \citenamefont {Bruun},\ and\ \citenamefont
  {Arlt}}]{PRL.121.173403}%
  \BibitemOpen
  \bibfield  {author} {\bibinfo {author} {\bibfnamefont {N.~B.}\ \bibnamefont
  {J\o{}rgensen}}, \bibinfo {author} {\bibfnamefont {G.~M.}\ \bibnamefont
  {Bruun}},\ and\ \bibinfo {author} {\bibfnamefont {J.~J.}\ \bibnamefont
  {Arlt}},\ }\bibfield  {title} {\bibinfo {title} {Dilute fluid governed by
  quantum fluctuations},\ }\href
  {https://doi.org/10.1103/PhysRevLett.121.173403} {\bibfield  {journal}
  {\bibinfo  {journal} {Phys. Rev. Lett.}\ }\textbf {\bibinfo {volume} {121}},\
  \bibinfo {pages} {173403} (\bibinfo {year} {2018})}\BibitemShut {NoStop}%
\bibitem [{\citenamefont {Cavicchioli}\ \emph {et~al.}(2025)\citenamefont
  {Cavicchioli}, \citenamefont {Fort}, \citenamefont {Ancilotto}, \citenamefont
  {Modugno}, \citenamefont {Minardi},\ and\ \citenamefont
  {Burchianti}}]{PRL.134.093401}%
  \BibitemOpen
  \bibfield  {author} {\bibinfo {author} {\bibfnamefont {L.}~\bibnamefont
  {Cavicchioli}}, \bibinfo {author} {\bibfnamefont {C.}~\bibnamefont {Fort}},
  \bibinfo {author} {\bibfnamefont {F.}~\bibnamefont {Ancilotto}}, \bibinfo
  {author} {\bibfnamefont {M.}~\bibnamefont {Modugno}}, \bibinfo {author}
  {\bibfnamefont {F.}~\bibnamefont {Minardi}},\ and\ \bibinfo {author}
  {\bibfnamefont {A.}~\bibnamefont {Burchianti}},\ }\bibfield  {title}
  {\bibinfo {title} {Dynamical formation of multiple quantum droplets in a
  {Bose-Bose} mixture},\ }\href
  {https://doi.org/https://doi.org/10.1103/PhysRevLett.134.093401} {\bibfield
  {journal} {\bibinfo  {journal} {Phys. Rev. Lett.}\ }\textbf {\bibinfo
  {volume} {134}},\ \bibinfo {pages} {093401} (\bibinfo {year}
  {2025})}\BibitemShut {NoStop}%
\bibitem [{\citenamefont {Khaykovich}\ \emph {et~al.}(2002)\citenamefont
  {Khaykovich}, \citenamefont {Schreck}, \citenamefont {Ferrari}, \citenamefont
  {Bourdel}, \citenamefont {Cubizolles}, \citenamefont {Carr}, \citenamefont
  {Castin},\ and\ \citenamefont {Salomon}}]{Science296.1290}%
  \BibitemOpen
  \bibfield  {author} {\bibinfo {author} {\bibfnamefont {L.}~\bibnamefont
  {Khaykovich}}, \bibinfo {author} {\bibfnamefont {F.}~\bibnamefont {Schreck}},
  \bibinfo {author} {\bibfnamefont {G.}~\bibnamefont {Ferrari}}, \bibinfo
  {author} {\bibfnamefont {T.}~\bibnamefont {Bourdel}}, \bibinfo {author}
  {\bibfnamefont {J.}~\bibnamefont {Cubizolles}}, \bibinfo {author}
  {\bibfnamefont {L.~D.}\ \bibnamefont {Carr}}, \bibinfo {author}
  {\bibfnamefont {Y.}~\bibnamefont {Castin}},\ and\ \bibinfo {author}
  {\bibfnamefont {C.}~\bibnamefont {Salomon}},\ }\bibfield  {title} {\bibinfo
  {title} {Formation of a matter-wave bright soliton},\ }\href
  {https://doi.org/10.1126/science.1071021} {\bibfield  {journal} {\bibinfo
  {journal} {Science}\ }\textbf {\bibinfo {volume} {296}},\ \bibinfo {pages}
  {1290} (\bibinfo {year} {2002})}\BibitemShut {NoStop}%
\bibitem [{\citenamefont {Strecker}\ \emph {et~al.}(2002)\citenamefont
  {Strecker}, \citenamefont {Partridge}, \citenamefont {Truscott},\ and\
  \citenamefont {Hulet}}]{Nature417.150}%
  \BibitemOpen
  \bibfield  {author} {\bibinfo {author} {\bibfnamefont {K.~E.}\ \bibnamefont
  {Strecker}}, \bibinfo {author} {\bibfnamefont {G.~B.}\ \bibnamefont
  {Partridge}}, \bibinfo {author} {\bibfnamefont {A.~G.}\ \bibnamefont
  {Truscott}},\ and\ \bibinfo {author} {\bibfnamefont {R.~G.}\ \bibnamefont
  {Hulet}},\ }\bibfield  {title} {\bibinfo {title} {Formation and propagation
  of matter-wave soliton trains},\ }\href {https://doi.org/10.1038/nature747}
  {\bibfield  {journal} {\bibinfo  {journal} {Nature}\ }\textbf {\bibinfo
  {volume} {417}},\ \bibinfo {pages} {150} (\bibinfo {year}
  {2002})}\BibitemShut {NoStop}%
\bibitem [{\citenamefont {Nguyen}\ \emph {et~al.}(2014)\citenamefont {Nguyen},
  \citenamefont {Dyke}, \citenamefont {Luo}, \citenamefont {Malomed},\ and\
  \citenamefont {Hulet}}]{NP10.918}%
  \BibitemOpen
  \bibfield  {author} {\bibinfo {author} {\bibfnamefont {J.~H.~V.}\
  \bibnamefont {Nguyen}}, \bibinfo {author} {\bibfnamefont {P.}~\bibnamefont
  {Dyke}}, \bibinfo {author} {\bibfnamefont {D.}~\bibnamefont {Luo}}, \bibinfo
  {author} {\bibfnamefont {B.~A.}\ \bibnamefont {Malomed}},\ and\ \bibinfo
  {author} {\bibfnamefont {R.~G.}\ \bibnamefont {Hulet}},\ }\bibfield  {title}
  {\bibinfo {title} {Collisions of matter-wave solitons},\ }\href
  {https://doi.org/10.1038/NPHYS3135} {\bibfield  {journal} {\bibinfo
  {journal} {Nat. Phys.}\ }\textbf {\bibinfo {volume} {10}},\ \bibinfo {pages}
  {918} (\bibinfo {year} {2014})}\BibitemShut {NoStop}%
\bibitem [{\citenamefont {Di~Carli}\ \emph {et~al.}(2019)\citenamefont
  {Di~Carli}, \citenamefont {Colquhoun}, \citenamefont {Henderson},
  \citenamefont {Flannigan}, \citenamefont {Oppo}, \citenamefont {Daley},
  \citenamefont {Kuhr},\ and\ \citenamefont {Haller}}]{PRL.123.123602}%
  \BibitemOpen
  \bibfield  {author} {\bibinfo {author} {\bibfnamefont {A.}~\bibnamefont
  {Di~Carli}}, \bibinfo {author} {\bibfnamefont {C.~D.}\ \bibnamefont
  {Colquhoun}}, \bibinfo {author} {\bibfnamefont {G.}~\bibnamefont
  {Henderson}}, \bibinfo {author} {\bibfnamefont {S.}~\bibnamefont
  {Flannigan}}, \bibinfo {author} {\bibfnamefont {G.-L.}\ \bibnamefont {Oppo}},
  \bibinfo {author} {\bibfnamefont {A.~J.}\ \bibnamefont {Daley}}, \bibinfo
  {author} {\bibfnamefont {S.}~\bibnamefont {Kuhr}},\ and\ \bibinfo {author}
  {\bibfnamefont {E.}~\bibnamefont {Haller}},\ }\bibfield  {title} {\bibinfo
  {title} {Excitation modes of bright matter-wave solitons},\ }\href
  {https://doi.org/10.1103/PhysRevLett.123.123602} {\bibfield  {journal}
  {\bibinfo  {journal} {Phys. Rev. Lett.}\ }\textbf {\bibinfo {volume} {123}},\
  \bibinfo {pages} {123602} (\bibinfo {year} {2019})}\BibitemShut {NoStop}%
\bibitem [{\citenamefont {Kengne}\ \emph {et~al.}(2021)\citenamefont {Kengne},
  \citenamefont {Liu},\ and\ \citenamefont {Malomed}}]{PR.899.1}%
  \BibitemOpen
  \bibfield  {author} {\bibinfo {author} {\bibfnamefont {E.}~\bibnamefont
  {Kengne}}, \bibinfo {author} {\bibfnamefont {W.-M.}\ \bibnamefont {Liu}},\
  and\ \bibinfo {author} {\bibfnamefont {B.~A.}\ \bibnamefont {Malomed}},\
  }\bibfield  {title} {\bibinfo {title} {Spatiotemporal engineering of
  matter-wave solitons in {Bose-Einstein} condensates},\ }\href
  {https://doi.org/https://doi.org/10.1016/j.physrep.2020.11.001} {\bibfield
  {journal} {\bibinfo  {journal} {Phys. Rep.}\ }\textbf {\bibinfo {volume}
  {899}},\ \bibinfo {pages} {1} (\bibinfo {year} {2021})}\BibitemShut {NoStop}%
\bibitem [{\citenamefont {Mossman}\ \emph {et~al.}(2024)\citenamefont
  {Mossman}, \citenamefont {Katsimiga}, \citenamefont {Mistakidis},
  \citenamefont {Romero-Ros}, \citenamefont {Bersano}, \citenamefont
  {Schmelcher}, \citenamefont {Kevrekidis},\ and\ \citenamefont
  {Engels}}]{CP.7.163}%
  \BibitemOpen
  \bibfield  {author} {\bibinfo {author} {\bibfnamefont {S.~M.}\ \bibnamefont
  {Mossman}}, \bibinfo {author} {\bibfnamefont {G.~C.}\ \bibnamefont
  {Katsimiga}}, \bibinfo {author} {\bibfnamefont {S.~I.}\ \bibnamefont
  {Mistakidis}}, \bibinfo {author} {\bibfnamefont {A.}~\bibnamefont
  {Romero-Ros}}, \bibinfo {author} {\bibfnamefont {T.~M.}\ \bibnamefont
  {Bersano}}, \bibinfo {author} {\bibfnamefont {P.}~\bibnamefont {Schmelcher}},
  \bibinfo {author} {\bibfnamefont {P.~G.}\ \bibnamefont {Kevrekidis}},\ and\
  \bibinfo {author} {\bibfnamefont {P.}~\bibnamefont {Engels}},\ }\bibfield
  {title} {\bibinfo {title} {Observation of dense collisional soliton complexes
  in a two-component {Bose-Einstein} condensate},\ }\href
  {https://doi.org/10.1038/s42005-024-01659-w} {\bibfield  {journal} {\bibinfo
  {journal} {Commun. Phys.}\ }\textbf {\bibinfo {volume} {7}},\ \bibinfo
  {pages} {163} (\bibinfo {year} {2024})}\BibitemShut {NoStop}%
\bibitem [{\citenamefont {Cruickshank}\ \emph {et~al.}(2025)\citenamefont
  {Cruickshank}, \citenamefont {Lorenzi}, \citenamefont {La~Rooij},
  \citenamefont {Kerr}, \citenamefont {Hilker}, \citenamefont {Kuhr},
  \citenamefont {Salasnich},\ and\ \citenamefont {Haller}}]{PRL.135.263404}%
  \BibitemOpen
  \bibfield  {author} {\bibinfo {author} {\bibfnamefont {R.}~\bibnamefont
  {Cruickshank}}, \bibinfo {author} {\bibfnamefont {F.}~\bibnamefont
  {Lorenzi}}, \bibinfo {author} {\bibfnamefont {A.}~\bibnamefont {La~Rooij}},
  \bibinfo {author} {\bibfnamefont {E.~F.}\ \bibnamefont {Kerr}}, \bibinfo
  {author} {\bibfnamefont {T.}~\bibnamefont {Hilker}}, \bibinfo {author}
  {\bibfnamefont {S.}~\bibnamefont {Kuhr}}, \bibinfo {author} {\bibfnamefont
  {L.}~\bibnamefont {Salasnich}},\ and\ \bibinfo {author} {\bibfnamefont
  {E.}~\bibnamefont {Haller}},\ }\bibfield  {title} {\bibinfo {title}
  {Experimental observation of single- and multisite matter-wave solitons in an
  optical accordion lattice},\ }\href {https://doi.org/10.1103/sh72-wnmv}
  {\bibfield  {journal} {\bibinfo  {journal} {Phys. Rev. Lett.}\ }\textbf
  {\bibinfo {volume} {135}},\ \bibinfo {pages} {263404} (\bibinfo {year}
  {2025})}\BibitemShut {NoStop}%
\bibitem [{\citenamefont {Malomed}(2016)}]{EPJST.225.2507}%
  \BibitemOpen
  \bibfield  {author} {\bibinfo {author} {\bibfnamefont {B.~A.}\ \bibnamefont
  {Malomed}},\ }\bibfield  {title} {\bibinfo {title} {Multidimensional
  solitons: Well-established results and novel findings},\ }\href
  {https://doi.org/10.1140/epjst/e2016-60025-y} {\bibfield  {journal} {\bibinfo
   {journal} {Eur. Phys. J. Spec. Top.}\ }\textbf {\bibinfo {volume} {225}},\
  \bibinfo {pages} {2507} (\bibinfo {year} {2016})}\BibitemShut {NoStop}%
\bibitem [{\citenamefont {Kartashov}\ \emph {et~al.}(2019)\citenamefont
  {Kartashov}, \citenamefont {Astrakharchik}, \citenamefont {Malomed},\ and\
  \citenamefont {Torner}}]{NRP.1.185}%
  \BibitemOpen
  \bibfield  {author} {\bibinfo {author} {\bibfnamefont {Y.~V.}\ \bibnamefont
  {Kartashov}}, \bibinfo {author} {\bibfnamefont {G.~E.}\ \bibnamefont
  {Astrakharchik}}, \bibinfo {author} {\bibfnamefont {B.~A.}\ \bibnamefont
  {Malomed}},\ and\ \bibinfo {author} {\bibfnamefont {L.}~\bibnamefont
  {Torner}},\ }\bibfield  {title} {\bibinfo {title} {Frontiers in
  multidimensional self-trapping of nonlinear fields and matter},\ }\href
  {https://doi.org/10.1038/s42254-019-0025-7} {\bibfield  {journal} {\bibinfo
  {journal} {Nat. Rev. Phys.}\ }\textbf {\bibinfo {volume} {1}},\ \bibinfo
  {pages} {185} (\bibinfo {year} {2019})}\BibitemShut {NoStop}%
\bibitem [{\citenamefont {Saito}\ and\ \citenamefont
  {Ueda}(2003)}]{PRL.90.040403}%
  \BibitemOpen
  \bibfield  {author} {\bibinfo {author} {\bibfnamefont {H.}~\bibnamefont
  {Saito}}\ and\ \bibinfo {author} {\bibfnamefont {M.}~\bibnamefont {Ueda}},\
  }\bibfield  {title} {\bibinfo {title} {Dynamically stabilized bright solitons
  in a two-dimensional {Bose-Einstein} condensate},\ }\href
  {https://doi.org/https://doi.org/10.1103/PhysRevLett.90.040403} {\bibfield
  {journal} {\bibinfo  {journal} {Phys. Rev. Lett.}\ }\textbf {\bibinfo
  {volume} {90}},\ \bibinfo {pages} {040403} (\bibinfo {year}
  {2003})}\BibitemShut {NoStop}%
\bibitem [{\citenamefont {Burger}\ \emph {et~al.}(1999)\citenamefont {Burger},
  \citenamefont {Bongs}, \citenamefont {Dettmer}, \citenamefont {Ertmer},
  \citenamefont {Sengstock}, \citenamefont {Sanpera}, \citenamefont
  {Shlyapnikov},\ and\ \citenamefont {Lewenstein}}]{PRL.83.5198}%
  \BibitemOpen
  \bibfield  {author} {\bibinfo {author} {\bibfnamefont {S.}~\bibnamefont
  {Burger}}, \bibinfo {author} {\bibfnamefont {K.}~\bibnamefont {Bongs}},
  \bibinfo {author} {\bibfnamefont {S.}~\bibnamefont {Dettmer}}, \bibinfo
  {author} {\bibfnamefont {W.}~\bibnamefont {Ertmer}}, \bibinfo {author}
  {\bibfnamefont {K.}~\bibnamefont {Sengstock}}, \bibinfo {author}
  {\bibfnamefont {A.}~\bibnamefont {Sanpera}}, \bibinfo {author} {\bibfnamefont
  {G.~V.}\ \bibnamefont {Shlyapnikov}},\ and\ \bibinfo {author} {\bibfnamefont
  {M.}~\bibnamefont {Lewenstein}},\ }\bibfield  {title} {\bibinfo {title} {Dark
  solitons in {Bose-Einstein} condensates},\ }\href
  {https://doi.org/10.1103/PhysRevLett.83.5198} {\bibfield  {journal} {\bibinfo
   {journal} {Phys. Rev. Lett.}\ }\textbf {\bibinfo {volume} {83}},\ \bibinfo
  {pages} {5198} (\bibinfo {year} {1999})}\BibitemShut {NoStop}%
\bibitem [{\citenamefont {Denschlag}\ \emph {et~al.}(2000)\citenamefont
  {Denschlag}, \citenamefont {Simsarian}, \citenamefont {Feder}, \citenamefont
  {Clark}, \citenamefont {Collins}, \citenamefont {Cubizolles}, \citenamefont
  {Deng}, \citenamefont {Hagley}, \citenamefont {Helmerson}, \citenamefont
  {Reinhardt}, \citenamefont {Rolston}, \citenamefont {Schneider},\ and\
  \citenamefont {Phillips}}]{Science287.5450}%
  \BibitemOpen
  \bibfield  {author} {\bibinfo {author} {\bibfnamefont {J.}~\bibnamefont
  {Denschlag}}, \bibinfo {author} {\bibfnamefont {J.~E.}\ \bibnamefont
  {Simsarian}}, \bibinfo {author} {\bibfnamefont {D.~L.}\ \bibnamefont
  {Feder}}, \bibinfo {author} {\bibfnamefont {C.~W.}\ \bibnamefont {Clark}},
  \bibinfo {author} {\bibfnamefont {L.~A.}\ \bibnamefont {Collins}}, \bibinfo
  {author} {\bibfnamefont {J.}~\bibnamefont {Cubizolles}}, \bibinfo {author}
  {\bibfnamefont {L.}~\bibnamefont {Deng}}, \bibinfo {author} {\bibfnamefont
  {E.~W.}\ \bibnamefont {Hagley}}, \bibinfo {author} {\bibfnamefont
  {K.}~\bibnamefont {Helmerson}}, \bibinfo {author} {\bibfnamefont {W.~P.}\
  \bibnamefont {Reinhardt}}, \bibinfo {author} {\bibfnamefont {S.~L.}\
  \bibnamefont {Rolston}}, \bibinfo {author} {\bibfnamefont {B.~I.}\
  \bibnamefont {Schneider}},\ and\ \bibinfo {author} {\bibfnamefont {W.~D.}\
  \bibnamefont {Phillips}},\ }\bibfield  {title} {\bibinfo {title} {Generating
  solitons by phase engineering of a {Bose-Einstein} condensate},\ }\href
  {https://doi.org/10.1126/science.287.5450.97} {\bibfield  {journal} {\bibinfo
   {journal} {Science}\ }\textbf {\bibinfo {volume} {287}},\ \bibinfo {pages}
  {97} (\bibinfo {year} {2000})}\BibitemShut {NoStop}%
\bibitem [{\citenamefont {Becker}\ \emph {et~al.}(2008)\citenamefont {Becker},
  \citenamefont {Stellmer}, \citenamefont {Soltan-Panahi}, \citenamefont
  {D{\"o}rscher}, \citenamefont {Baumert}, \citenamefont {Richter},
  \citenamefont {Kronj{\"a}ger}, \citenamefont {Bongs},\ and\ \citenamefont
  {Sengstock}}]{NP.4.496}%
  \BibitemOpen
  \bibfield  {author} {\bibinfo {author} {\bibfnamefont {C.}~\bibnamefont
  {Becker}}, \bibinfo {author} {\bibfnamefont {S.}~\bibnamefont {Stellmer}},
  \bibinfo {author} {\bibfnamefont {P.}~\bibnamefont {Soltan-Panahi}}, \bibinfo
  {author} {\bibfnamefont {S.}~\bibnamefont {D{\"o}rscher}}, \bibinfo {author}
  {\bibfnamefont {M.}~\bibnamefont {Baumert}}, \bibinfo {author} {\bibfnamefont
  {E.-M.}\ \bibnamefont {Richter}}, \bibinfo {author} {\bibfnamefont
  {J.}~\bibnamefont {Kronj{\"a}ger}}, \bibinfo {author} {\bibfnamefont
  {K.}~\bibnamefont {Bongs}},\ and\ \bibinfo {author} {\bibfnamefont
  {K.}~\bibnamefont {Sengstock}},\ }\bibfield  {title} {\bibinfo {title}
  {Oscillations and interactions of dark and dark-bright solitons in
  {Bose-Einstein} condensates},\ }\href {https://doi.org/10.1038/nphys962}
  {\bibfield  {journal} {\bibinfo  {journal} {Nat. Phys.}\ }\textbf {\bibinfo
  {volume} {4}},\ \bibinfo {pages} {496} (\bibinfo {year} {2008})}\BibitemShut
  {NoStop}%
\bibitem [{\citenamefont {Frantzeskakis}(2010)}]{JPA.43.213001}%
  \BibitemOpen
  \bibfield  {author} {\bibinfo {author} {\bibfnamefont {D.~J.}\ \bibnamefont
  {Frantzeskakis}},\ }\bibfield  {title} {\bibinfo {title} {Dark solitons in
  atomic {Bose-Einstein} condensates: From theory to experiments},\ }\href
  {https://doi.org/10.1088/1751-8113/43/21/213001} {\bibfield  {journal}
  {\bibinfo  {journal} {J. Phys. A: Math. Theor.}\ }\textbf {\bibinfo {volume}
  {43}},\ \bibinfo {pages} {213001} (\bibinfo {year} {2010})}\BibitemShut
  {NoStop}%
\bibitem [{\citenamefont {Nguyen}\ \emph {et~al.}(2017)\citenamefont {Nguyen},
  \citenamefont {Luo},\ and\ \citenamefont {Hulet}}]{Science356.422}%
  \BibitemOpen
  \bibfield  {author} {\bibinfo {author} {\bibfnamefont {J.~H.~V.}\
  \bibnamefont {Nguyen}}, \bibinfo {author} {\bibfnamefont {D.}~\bibnamefont
  {Luo}},\ and\ \bibinfo {author} {\bibfnamefont {R.~G.}\ \bibnamefont
  {Hulet}},\ }\bibfield  {title} {\bibinfo {title} {Formation of matter-wave
  soliton trains by modulational instability},\ }\href
  {https://doi.org/10.1126/science.aal3220} {\bibfield  {journal} {\bibinfo
  {journal} {Science}\ }\textbf {\bibinfo {volume} {356}},\ \bibinfo {pages}
  {422} (\bibinfo {year} {2017})}\BibitemShut {NoStop}%
\bibitem [{\citenamefont {Sanz}\ \emph {et~al.}(2022)\citenamefont {Sanz},
  \citenamefont {Fr\"olian}, \citenamefont {Chisholm}, \citenamefont
  {Cabrera},\ and\ \citenamefont {Tarruell}}]{PRL.128.013201}%
  \BibitemOpen
  \bibfield  {author} {\bibinfo {author} {\bibfnamefont {J.}~\bibnamefont
  {Sanz}}, \bibinfo {author} {\bibfnamefont {A.}~\bibnamefont {Fr\"olian}},
  \bibinfo {author} {\bibfnamefont {C.~S.}\ \bibnamefont {Chisholm}}, \bibinfo
  {author} {\bibfnamefont {C.~R.}\ \bibnamefont {Cabrera}},\ and\ \bibinfo
  {author} {\bibfnamefont {L.}~\bibnamefont {Tarruell}},\ }\bibfield  {title}
  {\bibinfo {title} {Interaction control and bright solitons in coherently
  coupled {Bose-Einstein} condensates},\ }\href
  {https://doi.org/10.1103/PhysRevLett.128.013201} {\bibfield  {journal}
  {\bibinfo  {journal} {Phys. Rev. Lett.}\ }\textbf {\bibinfo {volume} {128}},\
  \bibinfo {pages} {013201} (\bibinfo {year} {2022})}\BibitemShut {NoStop}%
\bibitem [{\citenamefont {Chiao}\ \emph {et~al.}(1964)\citenamefont {Chiao},
  \citenamefont {Garmire},\ and\ \citenamefont {Townes}}]{PRL.13.479}%
  \BibitemOpen
  \bibfield  {author} {\bibinfo {author} {\bibfnamefont {R.~Y.}\ \bibnamefont
  {Chiao}}, \bibinfo {author} {\bibfnamefont {E.}~\bibnamefont {Garmire}},\
  and\ \bibinfo {author} {\bibfnamefont {C.~H.}\ \bibnamefont {Townes}},\
  }\bibfield  {title} {\bibinfo {title} {Self-trapping of optical beams},\
  }\href {https://doi.org/10.1103/PhysRevLett.13.479} {\bibfield  {journal}
  {\bibinfo  {journal} {Phys. Rev. Lett.}\ }\textbf {\bibinfo {volume} {13}},\
  \bibinfo {pages} {479} (\bibinfo {year} {1964})}\BibitemShut {NoStop}%
\bibitem [{\citenamefont {Chen}\ and\ \citenamefont
  {Hung}(2020)}]{PRL.125.250401}%
  \BibitemOpen
  \bibfield  {author} {\bibinfo {author} {\bibfnamefont {C.-A.}\ \bibnamefont
  {Chen}}\ and\ \bibinfo {author} {\bibfnamefont {C.-L.}\ \bibnamefont
  {Hung}},\ }\bibfield  {title} {\bibinfo {title} {Observation of universal
  quench dynamics and {Townes} soliton formation from modulational instability
  in two-dimensional {Bose} gases},\ }\href
  {https://doi.org/https://doi.org/10.1103/PhysRevLett.125.250401} {\bibfield
  {journal} {\bibinfo  {journal} {Phys. Rev. Lett.}\ }\textbf {\bibinfo
  {volume} {125}},\ \bibinfo {pages} {250401} (\bibinfo {year}
  {2020})}\BibitemShut {NoStop}%
\bibitem [{\citenamefont {Chen}\ and\ \citenamefont
  {Hung}(2021)}]{PRL.127.023604}%
  \BibitemOpen
  \bibfield  {author} {\bibinfo {author} {\bibfnamefont {C.-A.}\ \bibnamefont
  {Chen}}\ and\ \bibinfo {author} {\bibfnamefont {C.-L.}\ \bibnamefont
  {Hung}},\ }\bibfield  {title} {\bibinfo {title} {Observation of scale
  invariance in two-dimensional matter-wave {Townes} solitons},\ }\href
  {https://doi.org/https://doi.org/10.1103/PhysRevLett.127.023604} {\bibfield
  {journal} {\bibinfo  {journal} {Phys. Rev. Lett.}\ }\textbf {\bibinfo
  {volume} {127}},\ \bibinfo {pages} {023604} (\bibinfo {year}
  {2021})}\BibitemShut {NoStop}%
\bibitem [{\citenamefont {Bakkali-Hassani}\ \emph {et~al.}(2021)\citenamefont
  {Bakkali-Hassani}, \citenamefont {Maury}, \citenamefont {Zou}, \citenamefont
  {Le~Cerf}, \citenamefont {Saint-Jalm}, \citenamefont {Castilho},
  \citenamefont {Nascimbene}, \citenamefont {Dalibard},\ and\ \citenamefont
  {Beugnon}}]{PRL.127.023603}%
  \BibitemOpen
  \bibfield  {author} {\bibinfo {author} {\bibfnamefont {B.}~\bibnamefont
  {Bakkali-Hassani}}, \bibinfo {author} {\bibfnamefont {C.}~\bibnamefont
  {Maury}}, \bibinfo {author} {\bibfnamefont {Y.-Q.}\ \bibnamefont {Zou}},
  \bibinfo {author} {\bibfnamefont {{\'E}.}~\bibnamefont {Le~Cerf}}, \bibinfo
  {author} {\bibfnamefont {R.}~\bibnamefont {Saint-Jalm}}, \bibinfo {author}
  {\bibfnamefont {P.~C.~M.}\ \bibnamefont {Castilho}}, \bibinfo {author}
  {\bibfnamefont {S.}~\bibnamefont {Nascimbene}}, \bibinfo {author}
  {\bibfnamefont {J.}~\bibnamefont {Dalibard}},\ and\ \bibinfo {author}
  {\bibfnamefont {J.}~\bibnamefont {Beugnon}},\ }\bibfield  {title} {\bibinfo
  {title} {Realization of a {Townes} soliton in a two-component planar {Bose}
  gas},\ }\href
  {https://doi.org/https://doi.org/10.1103/PhysRevLett.127.023603} {\bibfield
  {journal} {\bibinfo  {journal} {Phys. Rev. Lett.}\ }\textbf {\bibinfo
  {volume} {127}},\ \bibinfo {pages} {023603} (\bibinfo {year}
  {2021})}\BibitemShut {NoStop}%
\bibitem [{\citenamefont {Astrakharchik}\ and\ \citenamefont
  {Malomed}(2018)}]{PRA.98.013631}%
  \BibitemOpen
  \bibfield  {author} {\bibinfo {author} {\bibfnamefont {G.~E.}\ \bibnamefont
  {Astrakharchik}}\ and\ \bibinfo {author} {\bibfnamefont {B.~A.}\ \bibnamefont
  {Malomed}},\ }\bibfield  {title} {\bibinfo {title} {Dynamics of
  one-dimensional quantum droplets},\ }\href
  {https://doi.org/10.1103/PhysRevA.98.013631} {\bibfield  {journal} {\bibinfo
  {journal} {Phys. Rev. A}\ }\textbf {\bibinfo {volume} {98}},\ \bibinfo
  {pages} {013631} (\bibinfo {year} {2018})}\BibitemShut {NoStop}%
\bibitem [{\citenamefont {Schubert}\ \emph {et~al.}(2026)\citenamefont
  {Schubert}, \citenamefont {Bland}, \citenamefont {Mark}, \citenamefont
  {Ferlaino},\ and\ \citenamefont {Reimann}}]{PRR.8.023090}%
  \BibitemOpen
  \bibfield  {author} {\bibinfo {author} {\bibfnamefont {M.}~\bibnamefont
  {Schubert}}, \bibinfo {author} {\bibfnamefont {T.}~\bibnamefont {Bland}},
  \bibinfo {author} {\bibfnamefont {M.~J.}\ \bibnamefont {Mark}}, \bibinfo
  {author} {\bibfnamefont {F.}~\bibnamefont {Ferlaino}},\ and\ \bibinfo
  {author} {\bibfnamefont {S.~M.}\ \bibnamefont {Reimann}},\ }\bibfield
  {title} {\bibinfo {title} {Soliton-to-droplet crossover in a dipolar {Bose}
  gas in one and two dimensions},\ }\href
  {https://doi.org/https://doi.org/10.1103/lcd6-qz83} {\bibfield  {journal}
  {\bibinfo  {journal} {Phys. Rev. Res.}\ }\textbf {\bibinfo {volume} {8}},\
  \bibinfo {pages} {023090} (\bibinfo {year} {2026})}\BibitemShut {NoStop}%
\bibitem [{\citenamefont {Ferrier-Barbut}(2019)}]{PT.72.46}%
  \BibitemOpen
  \bibfield  {author} {\bibinfo {author} {\bibfnamefont {I.}~\bibnamefont
  {Ferrier-Barbut}},\ }\bibfield  {title} {\bibinfo {title} {Ultradilute
  quantum droplets},\ }\href {https://doi.org/10.1063/PT.3.4184} {\bibfield
  {journal} {\bibinfo  {journal} {Phys. Today}\ }\textbf {\bibinfo {volume}
  {72}},\ \bibinfo {pages} {46} (\bibinfo {year} {2019})}\BibitemShut {NoStop}%
\bibitem [{\citenamefont {Wang}\ \emph {et~al.}(2022)\citenamefont {Wang},
  \citenamefont {Wen}, \citenamefont {Liu}, \citenamefont {Li}, \citenamefont
  {Wang}, \citenamefont {Huang}, \citenamefont {Chen}, \citenamefont {Han},
  \citenamefont {Meng},\ and\ \citenamefont {Zhang}}]{CPB.31.103401}%
  \BibitemOpen
  \bibfield  {author} {\bibinfo {author} {\bibfnamefont {L.}~\bibnamefont
  {Wang}}, \bibinfo {author} {\bibfnamefont {K.}~\bibnamefont {Wen}}, \bibinfo
  {author} {\bibfnamefont {F.}~\bibnamefont {Liu}}, \bibinfo {author}
  {\bibfnamefont {Y.}~\bibnamefont {Li}}, \bibinfo {author} {\bibfnamefont
  {P.}~\bibnamefont {Wang}}, \bibinfo {author} {\bibfnamefont {L.}~\bibnamefont
  {Huang}}, \bibinfo {author} {\bibfnamefont {L.}~\bibnamefont {Chen}},
  \bibinfo {author} {\bibfnamefont {W.}~\bibnamefont {Han}}, \bibinfo {author}
  {\bibfnamefont {Z.}~\bibnamefont {Meng}},\ and\ \bibinfo {author}
  {\bibfnamefont {J.}~\bibnamefont {Zhang}},\ }\bibfield  {title} {\bibinfo
  {title} {Experimental realization of two-dimensional single-layer ultracold
  gases of $^{87}\mathrm{Rb}$ in an accordion lattice},\ }\href
  {https://doi.org/10.1088/1674-1056/ac873c} {\bibfield  {journal} {\bibinfo
  {journal} {Chin. Phys. B}\ }\textbf {\bibinfo {volume} {31}},\ \bibinfo
  {pages} {103401} (\bibinfo {year} {2022})}\BibitemShut {NoStop}%
\bibitem [{SM()}]{SM}%
  \BibitemOpen
  \href@noop {} {\bibinfo {title} {See supplemental material about the
  experimental details and theoretical models we used. {It} includes
  {Refs}.~\cite{PRL.120.235301, Science359.301, PRL.120.135301, PRA.98.051603,
  PRL.84.2551, PRA.106.013320, PRL.115.155302, PRL.117.100401,
  Nonlinearity21.R139, PRL.127.023603, PRL.13.479, OL.39.3429}.}}\BibitemShut
  {Stop}%
\bibitem [{\citenamefont {Petrov}\ \emph {et~al.}(2000)\citenamefont {Petrov},
  \citenamefont {Holzmann},\ and\ \citenamefont {Shlyapnikov}}]{PRL.84.2551}%
  \BibitemOpen
  \bibfield  {author} {\bibinfo {author} {\bibfnamefont {D.~S.}\ \bibnamefont
  {Petrov}}, \bibinfo {author} {\bibfnamefont {M.}~\bibnamefont {Holzmann}},\
  and\ \bibinfo {author} {\bibfnamefont {G.~V.}\ \bibnamefont {Shlyapnikov}},\
  }\bibfield  {title} {\bibinfo {title} {{Bose-Einstein} condensation in
  quasi-{2D} trapped gases},\ }\href
  {https://doi.org/10.1103/PhysRevLett.84.2551} {\bibfield  {journal} {\bibinfo
   {journal} {Phys. Rev. Lett.}\ }\textbf {\bibinfo {volume} {84}},\ \bibinfo
  {pages} {2551} (\bibinfo {year} {2000})}\BibitemShut {NoStop}%
\bibitem [{\citenamefont {Zin}\ \emph {et~al.}(2022)\citenamefont {Zin},
  \citenamefont {Pylak},\ and\ \citenamefont {Gajda}}]{PRA.106.013320}%
  \BibitemOpen
  \bibfield  {author} {\bibinfo {author} {\bibfnamefont {P.}~\bibnamefont
  {Zin}}, \bibinfo {author} {\bibfnamefont {M.}~\bibnamefont {Pylak}},\ and\
  \bibinfo {author} {\bibfnamefont {M.}~\bibnamefont {Gajda}},\ }\bibfield
  {title} {\bibinfo {title} {Self-consistent description of {Bose-Bose}
  droplets: Harmonically trapped quasi-two-dimensional droplets},\ }\href
  {https://doi.org/10.1103/PhysRevA.106.013320} {\bibfield  {journal} {\bibinfo
   {journal} {Phys. Rev. A}\ }\textbf {\bibinfo {volume} {106}},\ \bibinfo
  {pages} {013320} (\bibinfo {year} {2022})}\BibitemShut {NoStop}%
\bibitem [{\citenamefont {Carretero-Gonz{\'a}lez}\ \emph
  {et~al.}(2008)\citenamefont {Carretero-Gonz{\'a}lez}, \citenamefont
  {Frantzeskakis},\ and\ \citenamefont {Kevrekidis}}]{Nonlinearity21.R139}%
  \BibitemOpen
  \bibfield  {author} {\bibinfo {author} {\bibfnamefont {R.}~\bibnamefont
  {Carretero-Gonz{\'a}lez}}, \bibinfo {author} {\bibfnamefont {D.~J.}\
  \bibnamefont {Frantzeskakis}},\ and\ \bibinfo {author} {\bibfnamefont
  {P.~G.}\ \bibnamefont {Kevrekidis}},\ }\bibfield  {title} {\bibinfo {title}
  {Nonlinear waves in {Bose-Einstein} condensates: Physical relevance and
  mathematical techniques},\ }\href
  {https://doi.org/10.1088/0951-7715/21/7/R01} {\bibfield  {journal} {\bibinfo
  {journal} {Nonlinearity}\ }\textbf {\bibinfo {volume} {21}},\ \bibinfo
  {pages} {R139} (\bibinfo {year} {2008})}\BibitemShut {NoStop}%
\bibitem [{\citenamefont {Lushnikov}\ and\ \citenamefont
  {Vladimirova}(2014)}]{OL.39.3429}%
  \BibitemOpen
  \bibfield  {author} {\bibinfo {author} {\bibfnamefont {P.~M.}\ \bibnamefont
  {Lushnikov}}\ and\ \bibinfo {author} {\bibfnamefont {N.}~\bibnamefont
  {Vladimirova}},\ }\bibfield  {title} {\bibinfo {title} {Nonlinear combining
  of laser beams},\ }\href {https://doi.org/10.1364/OL.39.003429} {\bibfield
  {journal} {\bibinfo  {journal} {Opt. Lett.}\ }\textbf {\bibinfo {volume}
  {39}},\ \bibinfo {pages} {3429} (\bibinfo {year} {2014})}\BibitemShut
  {NoStop}%
\bibitem [{\citenamefont {Flynn}\ \emph {et~al.}(2023)\citenamefont {Flynn},
  \citenamefont {Parisi}, \citenamefont {Billam},\ and\ \citenamefont
  {Parker}}]{PRR.5.033167}%
  \BibitemOpen
  \bibfield  {author} {\bibinfo {author} {\bibfnamefont {T.~A.}\ \bibnamefont
  {Flynn}}, \bibinfo {author} {\bibfnamefont {L.}~\bibnamefont {Parisi}},
  \bibinfo {author} {\bibfnamefont {T.~P.}\ \bibnamefont {Billam}},\ and\
  \bibinfo {author} {\bibfnamefont {N.~G.}\ \bibnamefont {Parker}},\ }\bibfield
   {title} {\bibinfo {title} {Quantum droplets in imbalanced atomic mixtures},\
  }\href {https://doi.org/10.1103/PhysRevResearch.5.033167} {\bibfield
  {journal} {\bibinfo  {journal} {Phys. Rev. Res.}\ }\textbf {\bibinfo {volume}
  {5}},\ \bibinfo {pages} {033167} (\bibinfo {year} {2023})}\BibitemShut
  {NoStop}%
\bibitem [{\citenamefont {Flynn}\ \emph {et~al.}(2024)\citenamefont {Flynn},
  \citenamefont {Keepfer}, \citenamefont {Parker},\ and\ \citenamefont
  {Billam}}]{PRR.6.013209}%
  \BibitemOpen
  \bibfield  {author} {\bibinfo {author} {\bibfnamefont {T.~A.}\ \bibnamefont
  {Flynn}}, \bibinfo {author} {\bibfnamefont {N.~A.}\ \bibnamefont {Keepfer}},
  \bibinfo {author} {\bibfnamefont {N.~G.}\ \bibnamefont {Parker}},\ and\
  \bibinfo {author} {\bibfnamefont {T.~P.}\ \bibnamefont {Billam}},\ }\bibfield
   {title} {\bibinfo {title} {Harmonically trapped imbalanced quantum
  droplets},\ }\href {https://doi.org/10.1103/PhysRevResearch.6.013209}
  {\bibfield  {journal} {\bibinfo  {journal} {Phys. Rev. Res.}\ }\textbf
  {\bibinfo {volume} {6}},\ \bibinfo {pages} {013209} (\bibinfo {year}
  {2024})}\BibitemShut {NoStop}%
\bibitem [{\citenamefont {Englezos}\ \emph {et~al.}(2024)\citenamefont
  {Englezos}, \citenamefont {Schmelcher},\ and\ \citenamefont
  {Mistakidis}}]{PRA.110.023324}%
  \BibitemOpen
  \bibfield  {author} {\bibinfo {author} {\bibfnamefont {I.~A.}\ \bibnamefont
  {Englezos}}, \bibinfo {author} {\bibfnamefont {P.}~\bibnamefont
  {Schmelcher}},\ and\ \bibinfo {author} {\bibfnamefont {S.~I.}\ \bibnamefont
  {Mistakidis}},\ }\bibfield  {title} {\bibinfo {title} {Particle-imbalanced
  weakly interacting quantum droplets in one dimension},\ }\href
  {https://doi.org/10.1103/PhysRevA.110.023324} {\bibfield  {journal} {\bibinfo
   {journal} {Phys. Rev. A}\ }\textbf {\bibinfo {volume} {110}},\ \bibinfo
  {pages} {023324} (\bibinfo {year} {2024})}\BibitemShut {NoStop}%
\bibitem [{\citenamefont {Hu}\ \emph {et~al.}(2025)\citenamefont {Hu},
  \citenamefont {Wang}, \citenamefont {Pu},\ and\ \citenamefont
  {Liu}}]{PRA.111.023309}%
  \BibitemOpen
  \bibfield  {author} {\bibinfo {author} {\bibfnamefont {H.}~\bibnamefont
  {Hu}}, \bibinfo {author} {\bibfnamefont {J.}~\bibnamefont {Wang}}, \bibinfo
  {author} {\bibfnamefont {H.}~\bibnamefont {Pu}},\ and\ \bibinfo {author}
  {\bibfnamefont {X.-J.}\ \bibnamefont {Liu}},\ }\bibfield  {title} {\bibinfo
  {title} {Breakdown of the single-mode description of ultradilute quantum
  droplets in binary {Bose} mixtures: A perspective from a microscopic bosonic
  pairing theory},\ }\href {https://doi.org/10.1103/PhysRevA.111.023309}
  {\bibfield  {journal} {\bibinfo  {journal} {Phys. Rev. A}\ }\textbf {\bibinfo
  {volume} {111}},\ \bibinfo {pages} {023309} (\bibinfo {year}
  {2025})}\BibitemShut {NoStop}%
\bibitem [{\citenamefont {Luo}\ \emph {et~al.}()\citenamefont {Luo},
  \citenamefont {Zhang},\ and\ \citenamefont {Cui}}]{arXiv:2609.28913}%
  \BibitemOpen
  \bibfield  {author} {\bibinfo {author} {\bibfnamefont {T.}~\bibnamefont
  {Luo}}, \bibinfo {author} {\bibfnamefont {J.}~\bibnamefont {Zhang}},\ and\
  \bibinfo {author} {\bibfnamefont {X.}~\bibnamefont {Cui}},\ }\href@noop {}
  {\bibinfo {title} {Moir\'e droplet of ultracold {Bose} gases in a
  twisted-bilayer optical lattice}},\ \Eprint
  {https://arxiv.org/abs/2609.28913} {arXiv:2609.28913} \BibitemShut {NoStop}%
\bibitem [{\citenamefont {Meng}\ \emph {et~al.}(2023)\citenamefont {Meng},
  \citenamefont {Wang}, \citenamefont {Han}, \citenamefont {Liu}, \citenamefont
  {Wen}, \citenamefont {Gao}, \citenamefont {Wang}, \citenamefont {Chin},\ and\
  \citenamefont {Zhang}}]{Nature615.231}%
  \BibitemOpen
  \bibfield  {author} {\bibinfo {author} {\bibfnamefont {Z.}~\bibnamefont
  {Meng}}, \bibinfo {author} {\bibfnamefont {L.}~\bibnamefont {Wang}}, \bibinfo
  {author} {\bibfnamefont {W.}~\bibnamefont {Han}}, \bibinfo {author}
  {\bibfnamefont {F.}~\bibnamefont {Liu}}, \bibinfo {author} {\bibfnamefont
  {K.}~\bibnamefont {Wen}}, \bibinfo {author} {\bibfnamefont {C.}~\bibnamefont
  {Gao}}, \bibinfo {author} {\bibfnamefont {P.}~\bibnamefont {Wang}}, \bibinfo
  {author} {\bibfnamefont {C.}~\bibnamefont {Chin}},\ and\ \bibinfo {author}
  {\bibfnamefont {J.}~\bibnamefont {Zhang}},\ }\bibfield  {title} {\bibinfo
  {title} {Atomic {Bose-Einstein} condensate in twisted-bilayer optical
  lattices},\ }\href {https://doi.org/10.1038/s41586-023-05695-4} {\bibfield
  {journal} {\bibinfo  {journal} {Nature}\ }\textbf {\bibinfo {volume} {615}},\
  \bibinfo {pages} {231} (\bibinfo {year} {2023})}\BibitemShut {NoStop}%
\end{thebibliography}
\end{document}

\title{Supplemental Material for: Two-Dimensional Quantum Droplets in a Mixture of Bose-Einstein Condensates}

\author{Yue Zhang}
\affiliation{State Key Laboratory of Quantum Optics Technologies and Devices, Institute of Opto-electronics, \\
\makebox[0.8\linewidth][c]{Collaborative Innovation Center of Extreme Optics, Shanxi University, Taiyuan, Shanxi 030006, China}}

\author{Liangchao Chen}
\email[Contact author: ]{chenlchao87@sxu.edu.cn}
\affiliation{State Key Laboratory of Quantum Optics Technologies and Devices, Institute of Opto-electronics, \\
\makebox[0.8\linewidth][c]{Collaborative Innovation Center of Extreme Optics, Shanxi University, Taiyuan, Shanxi 030006, China}}
\affiliation{Hefei National Laboratory, Hefei, Anhui 230088, China}

\author{Wei Han}
\affiliation{State Key Laboratory of Quantum Optics Technologies and Devices, Institute of Opto-electronics, \\
\makebox[0.8\linewidth][c]{Collaborative Innovation Center of Extreme Optics, Shanxi University, Taiyuan, Shanxi 030006, China}}
\affiliation{Hefei National Laboratory, Hefei, Anhui 230088, China}

\author{Linshan Gong}
\affiliation{State Key Laboratory of Quantum Optics Technologies and Devices, Institute of Opto-electronics, \\
\makebox[0.8\linewidth][c]{Collaborative Innovation Center of Extreme Optics, Shanxi University, Taiyuan, Shanxi 030006, China}}

\author{Yazhou Wang}
\affiliation{State Key Laboratory of Quantum Optics Technologies and Devices, Institute of Opto-electronics, \\
\makebox[0.8\linewidth][c]{Collaborative Innovation Center of Extreme Optics, Shanxi University, Taiyuan, Shanxi 030006, China}}

\author{Zekui Wang}
\affiliation{State Key Laboratory of Quantum Optics Technologies and Devices, Institute of Opto-electronics, \\
\makebox[0.8\linewidth][c]{Collaborative Innovation Center of Extreme Optics, Shanxi University, Taiyuan, Shanxi 030006, China}}

\author{Xinhao Liang}
\affiliation{State Key Laboratory of Quantum Optics Technologies and Devices, Institute of Opto-electronics, \\
\makebox[0.8\linewidth][c]{Collaborative Innovation Center of Extreme Optics, Shanxi University, Taiyuan, Shanxi 030006, China}}

\author{Wenjie Sui}
\affiliation{State Key Laboratory of Quantum Optics Technologies and Devices, Institute of Opto-electronics, \\
\makebox[0.8\linewidth][c]{Collaborative Innovation Center of Extreme Optics, Shanxi University, Taiyuan, Shanxi 030006, China}}

\author{Pengjun Wang}
\affiliation{State Key Laboratory of Quantum Optics Technologies and Devices, Institute of Opto-electronics, \\
\makebox[0.8\linewidth][c]{Collaborative Innovation Center of Extreme Optics, Shanxi University, Taiyuan, Shanxi 030006, China}}
\affiliation{Hefei National Laboratory, Hefei, Anhui 230088, China}

\author{Lianghui Huang}
\affiliation{State Key Laboratory of Quantum Optics Technologies and Devices, Institute of Opto-electronics, \\
\makebox[0.8\linewidth][c]{Collaborative Innovation Center of Extreme Optics, Shanxi University, Taiyuan, Shanxi 030006, China}}
\affiliation{Hefei National Laboratory, Hefei, Anhui 230088, China}

\author{Zengming Meng}
\affiliation{State Key Laboratory of Quantum Optics Technologies and Devices, Institute of Opto-electronics, \\
\makebox[0.8\linewidth][c]{Collaborative Innovation Center of Extreme Optics, Shanxi University, Taiyuan, Shanxi 030006, China}}
\affiliation{Hefei National Laboratory, Hefei, Anhui 230088, China}

\author{Zhuxiong Ye}
\affiliation{State Key Laboratory of Quantum Optics Technologies and Devices, Institute of Opto-electronics, \\
\makebox[0.8\linewidth][c]{Collaborative Innovation Center of Extreme Optics, Shanxi University, Taiyuan, Shanxi 030006, China}}

\author{Jing Zhang}
\email[Contact author: ]{jzhang74@sxu.edu.cn}
\affiliation{State Key Laboratory of Quantum Optics Technologies and Devices, Institute of Opto-electronics, \\
\makebox[0.8\linewidth][c]{Collaborative Innovation Center of Extreme Optics, Shanxi University, Taiyuan, Shanxi 030006, China}}
\affiliation{Hefei National Laboratory, Hefei, Anhui 230088, China}

\maketitle

\section{I. Experimental details}

\subsection{A. Time-of-flight absorption imaging and in situ imaging}

In our experiments, self-bound quantum droplets and Townes solitons are detected using short-time-of-flight (short-TOF) absorption imaging. The imaging system incorporates an objective with a numerical aperture of $\mathrm{NA}=0.6$ and a magnification of 8.7. The probe light in the imaging system is resonant with the $4\,\mathrm{S}_{1/2}, F=2 \rightarrow 4\,\mathrm{P}_{3/2}, F=3$ transition of $^{39}\mathrm{K}$. To detect atoms in the $4\,\mathrm{S}_{1/2}, F=1$ level, a repumping light pulse is applied prior to the probe light pulse to transfer the atoms to $4\,\mathrm{S}_{1/2}, F=2$. This scheme provides high and nearly equal detection efficiency for both $\lvert\uparrow\rangle$ and $\lvert\downarrow\rangle$ states at low magnetic fields. For the absorption imaging, the accordion optical lattice (providing $\omega_z$) is first switched off, and the ultracold atomic cloud is allowed to expand for a duration of $T_1 = 2\, \mathrm{ms}$, during which the atomic density is sufficiently reduced. Subsequently, we turn off the Feshbach magnetic field while simultaneously turning on a separate auxiliary coil that provides a weak bias field for imaging. We then wait an additional $T_2 = 2\,\mathrm{ms}$ for the field to stabilize. Although the scattering lengths of both $\lvert\uparrow\rangle \equiv \lvert F=1,m_F=-1\rangle$ and $\lvert\downarrow\rangle \equiv \lvert 1,0\rangle$ states of $^{39}\mathrm{K}$ atoms are negative at such weak magnetic field, the reduced atomic density after $T_1$ renders negligible atom loss during the $T_2 = 2\,\mathrm{ms}$ interval.

%%%%%%%%%%%%%***Fig. 1***%%%%%%%%%%%%%%%%%%
 \begin{figure}[bp]
\centerline{\includegraphics[scale=1.0]{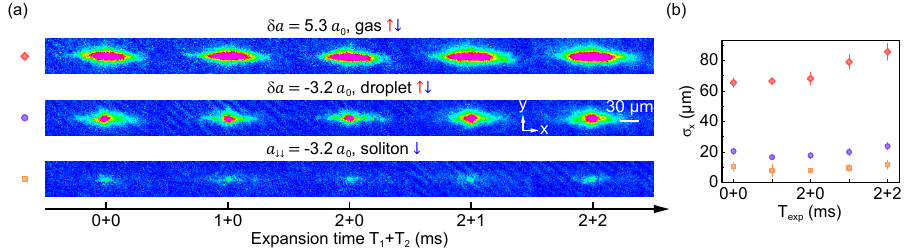}}
\caption{Representative absorption images of two-spin-mixture gaseous BECs, quantum droplets, and single-component Townes solitons. Before imaging, the ultracold atoms evolve for $30\,\mathrm{ms}$ in the accordion optical lattice without in-plane confinement. The TOF is varied during the absorption imaging sequence. Here, $T_1$ denotes the expansion time at the initial high magnetic field, while $T_2$ denotes the expansion time after the magnetic field is switched to and stabilized at a low field. (b) Cloud widths $\sigma_x$ extracted from (a), which remain nearly unchanged as the TOF increases.
}
\label{figS1}
\end{figure}

%%%%%%%%%%%%%****Fig. 1***%%%%%%%%%%%%%%%%%%

We can also perform in situ imaging of $^{39}\mathrm{K}$ atoms at high magnetic fields by setting $T_1 = 0\,\mathrm{ms}$ and $T_2 = 0\,\mathrm{ms}$, i.e., keeping the accordion optical lattice and the Feshbach magnetic field unchanged. The imaging system incorporates the same high-NA objective and a magnification of 39.6, yielding a calibrated spatial resolution of $1.5\,\mu\mathrm{m}$. At high magnetic field (around $56.8\,\mathrm{G}$), to efficiently detect atoms in the $\lvert\uparrow\rangle$  state, the probe light is optimized to be resonant with the $4\,\mathrm{S}_{1/2}, \lvert 2, 0 \rangle \rightarrow 4\,\mathrm{P}_{3/2}, \lvert 2, -1 \rangle $ transition of $^{39}\mathrm{K}$, while the repump light is optimized to be resonant with the $4\,\mathrm{S}_{1/2}, \lvert 1, -1 \rangle  \rightarrow 4\,\mathrm{P}_{3/2}, \lvert 2, 0 \rangle $ transition. For atoms in the $\lvert\downarrow\rangle$ state, however, no suitable level scheme is available in our optical setup to achieve efficient detection, resulting in greatly reduced detection efficiency for $\lvert\downarrow\rangle$ atoms. To image the $\lvert\downarrow\rangle$ state, we can apply an RF pulse before imaging to transfer all atoms to the $\lvert\uparrow\rangle$ state. However, it is difficult to efficiently detect both the $\lvert\uparrow\rangle$ and $\lvert\downarrow\rangle$ states simultaneously using in situ imaging at high magnetic fields.

Figure~\hyperref[figS1]{S1(a)} shows representative absorption images of gaseous BECs, droplets, and Townes solitons obtained by in situ imaging at high magnetic fields and short-TOF imaging at both high and low magnetic fields. All images are acquired using the same imaging system at a magnification of 39.6. The TOF is varied by incrementally increasing $T_1$ and $T_2$ before imaging. Figure~\hyperref[figS1]{S1(b)} presents the sizes of atomic clouds extracted from Fig.~\hyperref[figS1]{S1(a)}. The density profiles obtained after short TOF are found to closely resemble those measured by in situ imaging.

%%%%%%%%%%%%%***Fig. 2***%%%%%%%%%%%%%%%%%%
 \begin{figure}[t]
\centerline{\includegraphics[scale=1.0]{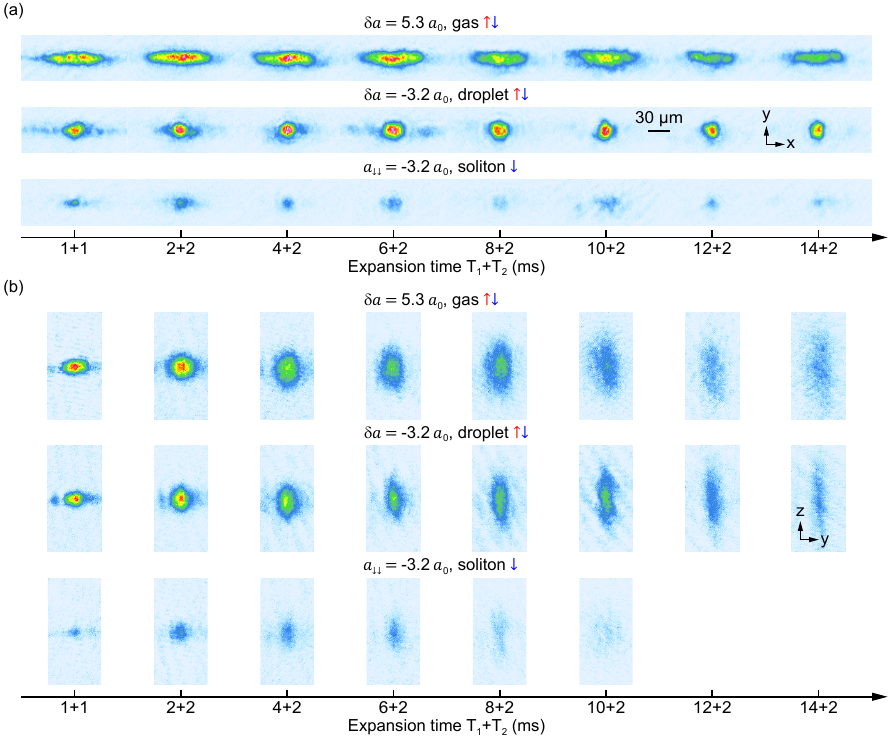}}
\caption{Expansion of the atomic clouds probed along (a) the vertical $z$ and (b) horizontal $x$ directions. The ultracold atoms are allowed to evolve for $30\,\mathrm{ms}$ in the accordion optical lattice without in-plane confinement before imaging. The atomic clouds expand predominantly along the vertical direction during TOF. Therefore, the horizontal density profiles obtained using short-TOF absorption imaging closely resemble those obtained using in situ imaging.}
\label{figS2}
\end{figure}

%%%%%%%%%%%%%****Fig. 2***%%%%%%%%%%%%%%%%%%

To further investigate the origin of the close resemblance between the atomic density profiles obtained after a short TOF and those measured in situ, we characterize in detail the expansion of ultracold $^{39}$K atoms following their release from the accordion optical lattice by probing along both the vertical and horizontal directions. Figure~\hyperref[figS2]{S2(a)} shows absorption images taken along the vertical ($z$) direction, in which we use the same high-NA objective with a magnification of 8.7 to improve the signal-to-noise ratio. Figure~\hyperref[figS2]{S2(b)} shows absorption images taken after the same TOF as in Fig.~\hyperref[figS2]{S2(a)}, along a horizontal direction at a small angle relative to the $x$ axis. As seen in Fig.~\hyperref[figS2]{S2}, after release from the accordion optical lattice, the gaseous BECs, quantum droplets, and Townes solitons expand rapidly along the $z$ direction due to the strong confinement in this direction, while their expansion in the horizontal ($xy$) plane remains negligible over short TOF. This strong anisotropy allows short-TOF absorption imaging along the vertical direction to yield density profiles that closely resemble those obtained in situ. For experimental convenience and simultaneous detection of both spin states, we therefore use short-TOF absorption imaging at a low magnetic field to probe the $^{39}$K atoms in the main text.

It should be noted that the gaseous BECs evolving in the quasi-2D geometry exhibit an elongated shape, as shown in Figs.~\hyperref[figS1]{S1(a)} and \hyperref[figS2]{S2(a)}. This elongation originates from a weak anisotropy in the spatial variation of the axial trapping frequency $\omega_z$ across the $x$-$y$ plane. The origin of this anisotropy can be understood from the geometry of the accordion optical lattice, which is formed by two focused laser beams, each with a circular waist of $\simeq 220\,\mu\mathrm{m}$, as shown in Fig.~1(a) of the main text. We define the $x$-direction along the bisector of the angle between the two laser beams and the $y$-direction normal to their common plane. Over the spatial extent of the atomic cloud,  the light intensity varies only weakly along the $x$-direction but considerably more rapidly along the $y$-direction. Consequently, the axial trapping frequency $\omega_z$ exhibits slightly different spatial dependences along the two directions.

\subsection{B. The lifetime of quantum droplets and Townes solitons}

\begin{figure}[ht]
\centerline{\includegraphics[scale=1.0]{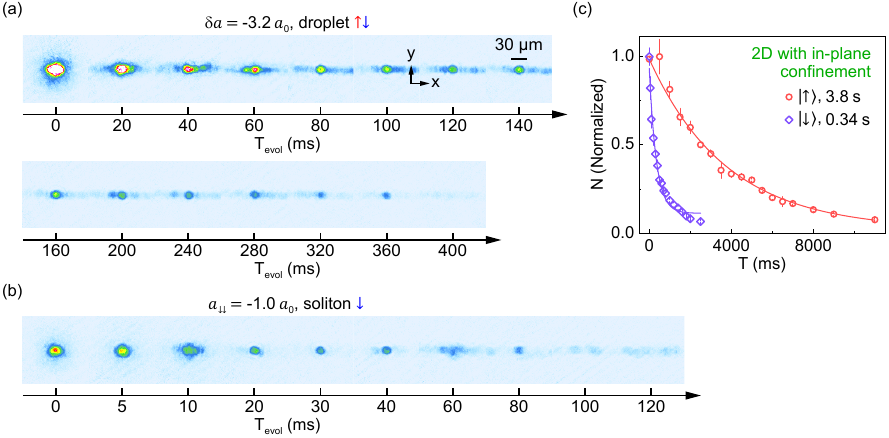}}
\caption{The complete evolution of quantum droplets (a) and Townes solitons (b) in the quasi‑2D geometry without in‑plane confinement. (c) The lifetimes of the single $\lvert\uparrow\rangle$ and $\lvert\downarrow\rangle$ BECs measured in a quasi‑2D geometry with in‑plane confinement. The magnetic field in (c) is set to $56.59\,\mathrm{G}$, corresponding to scattering lengths of $a_{\uparrow\uparrow}=33.7\,a_0$ and $a_{\downarrow\downarrow}=74.6\,a_0$ for the two spin states, respectively. Fitting the atom‑number ($N$) decay with exponential functions gives lifetimes of $3.8\,\mathrm{s}$ for the $\lvert\uparrow\rangle$ state and $0.34\,\mathrm{s}$ for the $\lvert\downarrow\rangle$ state.}
\label{figS3}
\end{figure}

We have examined the longest lifetimes of droplets and solitons under appropriate parameter conditions. Figures~\hyperref[figS3]{S3(a)} and \hyperref[figS3]{S3(b)} present the full temporal evolution of the droplets and solitons, respectively. While the soliton lifetime exceeds $50\,\mathrm{ms}$, the droplet lifetime exceeds $250\,\mathrm{ms}$. For comparison, previous experiments on quantum droplets formed from a two-spin mixture of $^{39}\mathrm{K}$ atoms reported typical lifetimes of $\sim\,7\,\mathrm{ms}$ in 3D free space~\cite{PRL.120.235301}, $\sim\,25\,\mathrm{ms}$ in a 3D geometry with weak confinement along the $z$-direction~\cite{Science359.301}, and $\sim\,50\,\mathrm{ms}$ in a 1D optical waveguide~\cite{PRL.120.135301}. The droplets prepared in our quasi-2D geometry exhibit lifetimes nearly an order of magnitude longer than those previously reported. Fig.~\hyperref[figS3]{S3(c)} shows that, under the quasi-2D geometry with in-plane confinement, the lifetimes of single-component BECs in the $\lvert\uparrow\rangle$ and $\lvert\downarrow\rangle$ states are $3.8\,\mathrm{s}$ and $0.34\,\mathrm{s}$, respectively.

%%%%%%%%%%%%%***Fig. 3***%%%%%%%%%%%%%%%%%%

%%%%%%%%%%%%%****Fig. 3***%%%%%%%%%%%%%%%%%%

\subsection{C. Population ratio of 2D quantum droplets}

\begin{figure}[htbp]
\centerline{\includegraphics[scale=1.0]{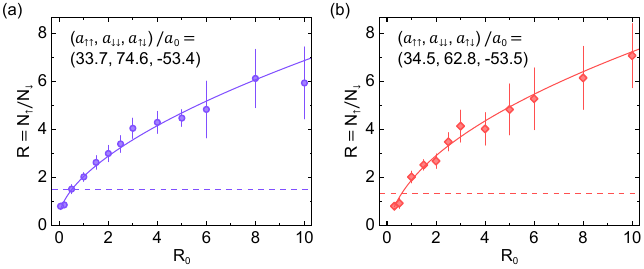}}
\caption{Range of the spin-population ratio for quantum droplets in quasi-2D geometry. The magnetic field is set to 56.59 G in (a) and 56.20 G in (b), corresponding to the scattering lengths $(a_{\uparrow\uparrow}, a_{\downarrow\downarrow}, a_{\uparrow\downarrow})/a_0=(33.7, 74.6, -53.4)$ and $(34.5, 62.8, -53.5)$, respectively. The solid lines are fits to the experimental data and serve as guides to the eye. The two dashed lines represent the theoretically optimal population ratios. In both (a) and (b), the mixtures are prepared with an initial population ratio $R_0$, and the population ratio $R$ is measured after $60\,\mathrm{ms}$ of evolution in a quasi-2D geometry without in-plane confinement. The two spin states are spatially separated during a longer TOF by the Stern–Gerlach gradient magnetic field.}
\label{figS4}
\end{figure}

Here, we provide a more detailed investigation of the spin-population ratio of 2D quantum droplets. In Figs.~\hyperref[figS4]{S4(a)} and~\hyperref[figS4]{S4(b)}, we measure the ratio $R = N_\uparrow/N_\downarrow$ for stable droplets across the full range accessible in our experiment. The magnetic field is tuned to $56.59\,\mathrm{G}$ and $56.20\,\mathrm{G}$, corresponding to the scattering lengths $(a_{\uparrow\uparrow}, a_{\downarrow\downarrow}, a_{\uparrow\downarrow})/a_0$ to be $(33.7, 74.6, -53.4)$ and $(34.5, 62.8, -53.5)$, respectively. The initial population ratio $R_0$ of the two-spin mixture is controlled by varying the duration of the RF pulse. After allowing the mixture to evolve for $60\,\mathrm{ms}$ in the quasi-2D geometry without in-plane confinement, stable droplets are formed, and then their spin-population ratios $R$ are measured. As shown in Figs.~\hyperref[figS4]{S4(a)} and~\hyperref[figS4]{S4(b)}, for both magnetic fields, $R$ can vary from approximately $0.8$ to $7$, substantially exceeding the stability range near the optimal population ratio predicted by earlier theoretical studies. We note that the error bars increase with $R$, as it becomes increasingly difficult to distinguish bound atoms in the droplet from unbound atoms within the majority component. Meanwhile, nearly all atoms in the minority component remain self-bound, providing clear evidence for the presence of the droplet.

%%%%%%%%%%%%%***Fig. 4***%%%%%%%%%%%%%%%%%%

%%%%%%%%%%%%%****Fig. 4***%%%%%%%%%%%%%%%%%%

%%%%%%%%%%%%%%%%%%%%%%%%%%%****II. Theory model*****%%%%%%%%%%%%%%%%%%%%%%%%%%%

\section{II. Theory model}

\subsection{A. Effective dimension of mean-field interactions and LHY correction}

For a single component gas in a quasi-2D geometry, when the axial ($z$-direction) harmonic confinement is much stronger than the low-energy in-plane dynamics of the atoms, we can employ the frozen-mode approximation. In this approximation, the axial wave function is frozen in the ground state of the harmonic trap, and the atomic wave function can then be written as
\begin{equation}
	\Psi(\mathbf{r}, z, t) = \psi(\mathbf{r}, t) \chi(z) e^{-i \omega_z t / 2},
	\label{eq:townes-factorization}
\end{equation}
in which, $\int dz\,|\chi|^2=1$, $\int d^2\mathbf{r}\,|\psi|^2=N$, with $N$ being the atom number, and $\omega_z$ the axial trapping frequency. Then the effective 2D coupling strength is given by
\begin{equation}
	g^{\mathrm{2D}}=g^{\mathrm{3D}}I_z,
	\label{eq:townes-g2d-general}
\end{equation}
with $g^{\mathrm{3D}}=4\pi\hbar^2 a/m$ and $I_z=\int dz |\chi(z)|^4$. Here, $a$ is the 3D scattering length. For the axial ground state of the harmonic trap, we have

\begin{equation}
	I_z=\frac{1}{\sqrt{2\pi}a_z},
	\label{eq:Iz}
\end{equation}
and therefore
\begin{equation}
	g^{\mathrm{2D}}=\frac{\sqrt{8\pi} \hbar^2 a}{ma_z},
	\label{eq:townes-harmonic-coupling}
\end{equation}
where $a_z=\sqrt{\hbar/(m\omega_z)}$ is the harmonic oscillator length, and can be used as a natural unit of length.

For the two-component mixture with optimal population ratio $N_{\uparrow}/N_{\downarrow}=\sqrt{a_{\downarrow\downarrow}/a_{\uparrow\uparrow}}$ in the 2D geometry, the residual 2D coupling constant is
\begin{equation}
g_{\mathrm{eff}}^{\mathrm{2D}}
=g_{\uparrow\downarrow}^{\mathrm{2D}}
+\sqrt{g_{\uparrow\uparrow}^{\mathrm{2D}}g_{\downarrow\downarrow}^{\mathrm{2D}}}
=\sqrt{8\pi}\hbar^2\delta a/(ma_z).
\end{equation}
Here, $g^{\mathrm{2D}}_{ij}=\sqrt{8\pi} \hbar^2 a_{ij} /(m a_z)$, and the subscripts $\{ i,j \}=\{\uparrow, \downarrow \}$ represent the two components $\lvert\uparrow\rangle$ and $\lvert\downarrow\rangle$ in the mixture, and $\delta a=a_{\uparrow\downarrow}+\sqrt{a_{\uparrow\uparrow}a_{\downarrow\downarrow}}$.

The mean-field condensate dynamics and the quantum fluctuations responsible for the Lee-Huang-Yang (LHY) correction are characterized by markedly different energy scales. Their effective dimensionalities must therefore be assessed separately by considering the soft and stiff Bogoliubov branches~\cite{PRA.98.051603, PRL.84.2551}. In the mean-field approximation, the low-energy condensate dynamics is governed by the soft branch with
\begin{equation}
	\eta_{\rm soft}
	=  \frac{	\max\!\left\{
		|g_{\rm eff}^{2D}|n_{\rm pk}, k_{\mathrm{B}}T
		\right\} }{\hbar\omega_z}.
	\label{eq:dimension-soft}
\end{equation}
When $\eta_{\rm soft}\ll1$, the condensates can be considered as quasi-2D. For our experimental parameters (see Table~\ref{tabS1}), the two-spin mixture in the droplet-phase interaction regime has an estimated temperature of $13(4)\,\mathrm{nK}$. Consequently, we have
\begin{equation}
\frac{k_{\mathrm{B}}T}{\hbar\omega_z} \simeq 7.4\times10^{-2}.
\end{equation}
Meanwhile, the mean-field interaction scale gives
\begin{equation}
	\frac{  |g_{\mathrm{eff}}^{\mathrm{2D}}|n_{\mathrm{pk}}  }{  \hbar\omega_z  }
	=\sqrt{8\pi} |\delta a| a_z n_{\mathrm{pk}}
	\simeq 3.5\times10^{-3}.
	\label{eq:dimension-soft-numbers}
\end{equation}
Both the thermal energy and residual mean-field interaction energy are much smaller than the axial energy-level spacing. Therefore, the condensate dynamic is effectively 2D at the mean-field level.

The dimensionality of the beyond-mean-field contribution is instead controlled by the stiff Bogoliubov branch, which gives the dominant contribution to the LHY energy near mean-field cancellation. The dimensionality can be determined by
\begin{equation}
	\chi_{\mathrm{stiff}}
	= \frac{g_{\uparrow\uparrow}^{\mathrm{3D}}n_{\uparrow}^{\mathrm{3D}}
	+  g_{\downarrow\downarrow}^{\mathrm{3D}}n_{\downarrow}^{\mathrm{3D}}
	}{  \hbar\omega_z  },
	\label{eq:dimension-stiff}
\end{equation}
where $g_{ij}^{\mathrm{3D}}=4\pi\hbar^2a_{ij}/m$, $n_{\uparrow}^{\mathrm{3D}}$ and $n_{\downarrow}^{\mathrm{3D}}$ are the atomic number densities. In the frozen-mode approximation, each component in the mixture has a peak atomic number density $n_i^{\mathrm{3D}}=n_i^{\mathrm{2D}}/(\sqrt{\pi}a_z)$. While the 3D local-density approximation requires $\chi_{\mathrm{stiff}}\gg1$~\cite{PRA.106.013320}, the LHY correction can be considered as quasi-2D for $\chi_{\mathrm{stiff}}\ll1$. With our experimental parameters, $\chi_{\mathrm{stiff}}\simeq  7.9\times10^{-2}\ll1$, and therefore the LHY correction can be considered as effective~2D.
\vspace{\baselineskip}

%%%%%%%%%%%%%%%%% Table S1 %%%%%%%%%%%%%%%%%

\begin{table}[htbp]
\vspace{-1 em}
\renewcommand{\arraystretch}{1.2}
\setlength{\belowcaptionskip}{1.5mm}
\caption{Representative experimental parameters of the two-component mixture in the quasi-2D geometry.}
\centering
\setlength{\tabcolsep}{4mm}
\begin{tabular}{rll}

\toprule
Quantity & Definition & Value \\
\midrule
atomic species                   &   $^{39}\mathrm{K}$        &  --  \\
atomic mass                       & $m$                                 & $38.9637\,\mathrm{u}$ \\
axial confinement                & $\omega_z/(2\pi)$          & $3.67\,\mathrm{kHz}$ \\
in-plane confinement          & $\omega_{x,y}/(2\pi)$    & $42\,\mathrm{Hz}$ \\
axial oscillator length          & $a_z=\sqrt{\hbar/(m\omega_z)}$      & $0.266\,\mu\mathrm{m}$ \\
scattering lengths               & $(a_{\uparrow\uparrow},a_{\downarrow\downarrow},a_{\uparrow\downarrow})/a_0$                    & $(33.7, 74.6, -53.4)$ \\
residual attraction               & $\delta a=a_{\uparrow\downarrow}+\sqrt{a_{\uparrow\uparrow}a_{\downarrow\downarrow}}$       & $-3.2\,a_0$ \\
optimal density ratio           & $R_{\mathrm{opt}}=\sqrt{a_{\downarrow\downarrow}/a_{\uparrow\uparrow}}$                                & $1.49$ \\
temperature                        & $T$                 & $13(4)\,\mathrm{nK}$ \\
atom number                      & $N$                 & $3.6\times10^4$ \\
radial width                         & $\sigma_\mathrm{r}$                 & $27\,\mu\mathrm{m}$ \\
peak total areal density      & $n_{\mathrm{pk}} = N/(\pi \sigma_\mathrm{r}^2)$                 & $16\,\mu\mathrm{m}^{-2}$ \\
\bottomrule

\end{tabular}
\vspace{0 em}
\label{tabS1}
\end{table}

%%%%%%%%%%%%%%%%% Table S1 %%%%%%%%%%%%%%%%%

%%%%%%%%%%%%%%%%%%%%%%%%%%%***B. 2D quantum droplets***%%%%%%%%%%%%%%%%%%%%%%%%%%%

\clearpage
\subsection{B. 2D quantum droplets}

The energy functional of the two-component mixture is given by

\begin{align}
	E=&\int \mathrm{d}^3 \mathbf{r} \Bigg\{
	\sum_{i=\uparrow,\downarrow}  \left[\dfrac{\hbar^2}{2m} |\nabla\psi_i|^2+V_\perp n_i\right]
	+\frac{g_{\uparrow\uparrow}}{2}n_\uparrow^2
	+\frac{g_{\downarrow\downarrow}}{2}n_\downarrow^2
	+g_{\uparrow\downarrow}n_\uparrow n_\downarrow
	+E_{\mathrm{LHY}}(n_\uparrow,n_\downarrow)
	\Bigg\},
	\label{eq:q2d-functional}
\end{align}
where $\psi_i$ are the wavefunctions of the two components with $n_i=|\psi_i|^2$ and $\int \mathrm{d}^3 \mathbf{r} n_i=N_i$. $V_\perp$ is the harmonic potential along the $z$-direction. In the droplet regime, minimization of the stiff spin-density mode locks the two components to the optimal density ratio~\cite{PRL.115.155302}
\begin{equation}
	R_{\mathrm{opt}}\equiv
	\frac{n_\uparrow}{n_\downarrow}
	= \sqrt{  \frac{a_{\downarrow\downarrow}}{a_{\uparrow\uparrow}}  },
	\qquad
	n=n_\uparrow+n_\downarrow.
	\label{eq:variational-optimal-ratio}
\end{equation}

\begin{figure}[htbp]
\centerline{\includegraphics[scale=1.0]{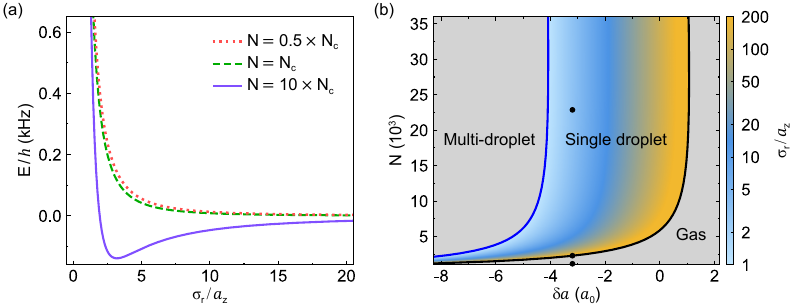}}
\caption{(a) Energy as a function of the condensate size $\sigma_\mathrm{r}$. The dotted, dashed, and solid lines correspond to atom number $N=0.5\times N_\mathrm{c}$, $N=N_\mathrm{c}$, and $N=10\times N_\mathrm{c}$, respectively, where $N_\mathrm{c}$ is the critical atom number for the formation of a stable droplet. (b) Theoretically predicted droplet phase diagram as a function of particle number $N$ and mean-field scattering length $\delta a$. The boundary colored by the black line is determined by $E_{\min}=0$, and the boundary colored by the blue line is determined by $\sigma_\mathrm{r}=a_z$. The color in the single droplet phase represents the size of the droplet. The three solid circles indicate the parameters corresponding to the energy functionals in (a).}
\label{figS5}
\end{figure}

Following the method presented in Ref.~\cite {PRL.117.100401}, we can reduce the two-component problem as an effective single-component with the dimensionless quasi-2D energy functional
\begin{equation}
	E[\phi] \equiv \int d^2 \mathbf{\rho}
	\left[ 	\frac{1}{2}|\tilde{\nabla}\phi|^2
	+
	\frac{g^2}{8\pi}n^2 \left( \ln\frac{n}{n_0}-1 \right)
	\right],
	\label{eq:effective-2d-energy}
\end{equation}
where the parameters $g$ and $n_0$ are defined as
\begin{equation}
	g=\frac{4\pi}{\ln(\epsilon_{\uparrow\downarrow}/\Delta)},
	\qquad
	n_0=e^{-3/2}\frac{\Delta}{|g|},
\end{equation}
with
\begin{align}
	\epsilon_{ij}
	&= \frac{B}{\pi} \exp\left(  \frac{\sqrt{2\pi}\, a_z}{a_{ij}}  \right),
	\qquad B \simeq 0.9,
	\\
	\Delta
        &=  \sqrt{  \epsilon_{\uparrow\downarrow} \sqrt{ \epsilon_{\uparrow\uparrow}\epsilon_{\downarrow\downarrow}  } }
	\exp\left[ -\frac{  \ln^2(\epsilon_{\uparrow\uparrow}/\epsilon_{\downarrow\downarrow})
	}{  4\ln\left(  \epsilon_{\uparrow\uparrow}\epsilon_{\downarrow\downarrow} / \epsilon_{\uparrow\downarrow}^{\, 2} \right)
	} \right].
\end{align}
By using a normalized Gaussian ansatz
\begin{equation}
	\phi
	=
	\dfrac{1}{\sqrt{\pi}\sigma}
	\exp\left(-\dfrac{\rho^2}{2\sigma^2}\right),
	\label{eq:2D gaussian ansatz}
\end{equation}
 and substituting Eq.~\eqref{eq:2D gaussian ansatz} into Eq.~\eqref{eq:effective-2d-energy}, we obtain the energy per particle
\begin{equation}
	E = \dfrac{1}{2\sigma^2}+\dfrac{g^2N}{16\pi^2\sigma^2}\left(\ln\left(\dfrac{N}{\pi\sigma^2n_0}\right)-\dfrac{3}{2}\right),
	\label{eq: 2d-droplet-energy}
\end{equation}
where the variational width $\sigma$ characterizes the size of the droplet. The energy as a function of droplet size is shown in Fig.~\hyperref[figS5]{S5(a)}. For a given interaction strength, we find a critical atom number. Below this critical value, the energy functional decreases monotonically with increasing cloud size, indicating that the atomic cloud tends to expand indefinitely. The system therefore remains in the gaseous phase in this parameter regime. Above the critical atom number, the energy functional develops a minimum at a finite cloud size, indicating that the system supports the existence of a self-bound quantum droplet.

%%%%%%%%%%%%%%%%% Fig. S5 %%%%%%%%%%%%%%%%%
%%%%%%%%%%%%%%%%% Fig. S5 %%%%%%%%%%%%%%%%%

We further examine the minimum of the energy functional over a broader range of interaction strengths and atom numbers, and obtain the droplet phase diagram, as shown in Fig.~\hyperref[figS5]{S5(b)}. The droplet phase is identified by a finite-size local minimum with $E_{\mathrm{min}}<0$, whereas the gas phase has no finite-size local minimum.  In addition, in the region where $\sigma<1$, i.e., where the condensate size $\sigma_\mathrm{r}<a_z$, the possible droplet size predicted in this region is much smaller than the initial size of the atomic cloud in the experiment, and this large mismatch may cause the cloud to split into multiple droplets.

%%%%%%%%%%%%%%%%%%%%%%%%%%%***C. Solitons in different dimensions***%%%%%%%%%%%%%%%%%%%%%%%%%%%
\subsection{C. Solitons in different dimensions}

Neglecting the LHY correction, attractive ultracold atoms can support soliton solutions in low-dimensional geometries, whereas attractive condensates in 3D free space undergo collapse~\cite{Nonlinearity21.R139}. This can be demonstrated from the energy functional in different spatial dimensions by solving the nonlinear Schrödinger equation

\begin{equation}
	-\frac{1}{2}\nabla^2\phi
	+g_{\mathrm{eff}}|\phi|^2\phi
	=\mu\phi .
	\label{eq:bright-soliton-nlse}
\end{equation}

\begin{figure}[h!]
\centerline{\includegraphics[scale=1]{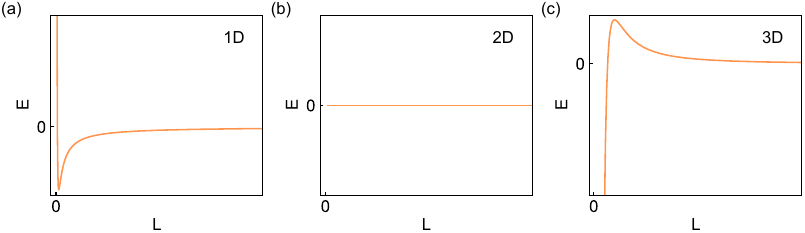}}
\caption{Energy as a function of atomic cloud size in different spatial dimensions. One- and two-dimensional systems support soliton solutions, whereas 3D systems do not. }
\label{figS6}
\end{figure}

Figure~\hyperref[figS6]{S6} shows the energy functional as a function of the wave-packet size for different dimensions. Here, the 1D soliton solution has a $\mathrm{sech}$-function form, while Gaussian ansatz wave functions are employed in two and three dimensions. In one dimension, the energy functional exhibits a minimum at a finite wave-packet size, indicating the existence of stable matter-wave solitons. In two dimensions, the energy functional near a critical atom number determined by the scattering length forms a horizontal line, reflecting the scale invariance of the system and allowing for the formation of marginally stable Townes solitons. In 3D free space, the energy functional decreases without bound as the wave-packet size approaches zero, indicating that the wave packet ultimately undergoes collapse.

%%%%%%%%%%%%%%%%% Fig. S6 %%%%%%%%%%%%%%%%%

%%%%%%%%%%%%%%%%% Fig. S6 %%%%%%%%%%%%%%%%%

\subsection{D. 2D Townes solitons}

%%%%%%%%%%%%%%%%% Table S2 %%%%%%%%%%%%%%%%%
\begin{table}[bph]
\vspace{-1 em}
\renewcommand{\arraystretch}{1.2}
\setlength{\belowcaptionskip}{1.5mm} %
\caption{Representative experimental parameters of the single-component BECs in the quasi-2D geometry. }
\centering
\setlength{\tabcolsep}{4mm}
\begin{tabular}{rll}

\toprule
Quantity                           & Definition                                                & Value \\
\midrule
atomic mass                    & $m$                                                        & $38.9637\,\mathrm{u}$ \\
axial confinement            &  $\omega_z/2\pi$                                    &  $3.67\,\mathrm{kHz}$  \\
in-plane confinement       &  $\omega_{x,y}/2\pi$                             &  $42\,\mathrm{Hz}$   \\
axial oscillator length       & $a_z=\sqrt{\hbar/(m\omega_z)}$           & $0.266\,\mu\mathrm{m}$ \\
scattering length              &  $a_{\downarrow\downarrow}$              &  $-3.2\,a_0$   \\
atom number                   & $N$                                                        & $2800$ \\
radial width                      & $\sigma_{\mathrm{r}}$                                                & $12\,\mu\mathrm{m}$ \\
peak areal density   &  $n_{\mathrm{pk}}=N/(\pi \sigma_{\mathrm{r}}^2)$                  &  $6\,\mu\mathrm m^{-2}$   \\
\bottomrule
\end{tabular}
\vspace{0 em}
\label{tabS2}
\end{table}
%%%%%%%%%%%%%%%%% Table S2 %%%%%%%%%%%%%%%%%

The relevant experimental parameters for 2D Townes solitons are summarized in Table~\ref{tabS2}. As in the two-component mixture, we identify the onset of the 2D regime by examining the soft Bogoliubov branch
\begin{equation}
	\eta_{\mathrm{soft}}
	= \frac{  |g^\mathrm{2D}| n_\mathrm{pk}  }  {  \hbar\omega_z  }
	= \sqrt{8\pi} |a_{\downarrow\downarrow}| a_z n_{\rm pk}.
	\label{eq:dimension-townes-number}
\end{equation}
According to the experimental parameters, we have $\eta_{\mathrm{soft}}= 1.4 \times10^{-3} \ll 1$, thus the system is well within the 2D regime.

\begin{figure}[H]
\centerline{\includegraphics[scale=1.0]{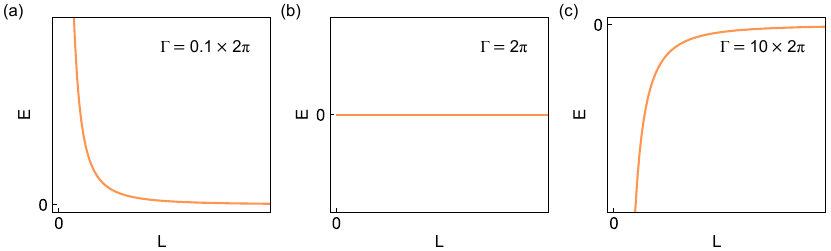}}
\caption{Energy as a function of the condensate width for different coupling strengths $\Gamma$ in a 2D system. The coupling strengths in (a), (b) and (c) are set to $\Gamma=0.1\times 2\pi$, $2\pi$, and $10\times 2\pi$, respectively.}
\label{figS7}
\end{figure}

The free evolution of 2D ultracold atoms is governed by
\begin{equation}
	i\hbar\partial_t\psi=
	\left[-\frac{\hbar^2}{2m}\nabla_{r}^2
	+g_{\downarrow\downarrow}^{2D}|\psi|^2\right]\psi.
	\label{eq:townes-physical-gpe}
\end{equation}
By setting $\bm{r}=\ell\bm{\rho}$, $t=m\ell^2\tau/\hbar$, and $\psi=\sqrt{N}\phi/\ell$, where $\ell$ is an arbitrary unit of length, Eq.~\eqref{eq:townes-physical-gpe} can be transformed into the dimensionless form
\begin{equation}
	i\partial_\tau\phi=-\frac12\nabla_\rho^2\phi-\Gamma|\phi|^2\phi,
	\label{eq:townes-dimensionless-gpe}
\end{equation}
with $\int d^2\bm{\rho}|\phi|^2=1$ and $\Gamma=4\pi N|a_{\downarrow\downarrow}|I_z$. The dimensionless energy per particle is
\begin{equation}
        E[\phi] = \int d^2\bm{\rho}
        \left[   \frac{1}{2} |\nabla_\rho \phi|^2
        - \frac{\Gamma}{2} |\phi|^4     \right].
        \label{eq:townes-energy}
\end{equation}
The 2D cubic equation~\eqref{eq:townes-dimensionless-gpe} is invariant under a simultaneous rescaling of space $\rho$, time $\tau$, and amplitude $\phi$ defined by $\rho=\lambda \tilde{\rho}$, $\tau=\lambda^2 \tilde{\tau}$, and $\phi=\lambda^{-1} \tilde{\phi}$. Thus, if $\phi(\bm\rho,\tau)$ is a solution of Eq.~\ref{eq:townes-dimensionless-gpe}, then so is $\tilde{\phi}(\bm\tilde{\rho},\tilde{\tau})$ with $\int d^2\tilde{\rho}\,|\tilde{\phi}|^2=1$. At the same time, the energy functional scales as $E[\tilde{\phi}]=\lambda^2 E[\phi]$~\cite{PRL.127.023603}. The scaling invariance implies that the energy functional has no finite-width minimum. Within a family of rescaled profiles, a positive-energy state approaches zero energy under dilation, whereas a negative-energy state is unbounded under compression.
This behavior can be illustrated by evaluating the energy of a normalized Gaussian ansatz as a function of the condensate width. Substituting Eq.~\eqref{eq:2D gaussian ansatz} into Eq.~\eqref{eq:townes-energy}, we obtain
\begin{equation}
	E(\sigma)
	=  \dfrac{1}{2\sigma^2}\left( 1-\frac{\Gamma}{2\pi} \right).
	\label{eq:2D_attractive_energy_analytic}
\end{equation}
Figure~\hyperref[figS7]{S7} shows the energy as a function of the condensate width for different coupling strengths. For $\Gamma<2\pi$, the energy is positive and decreases monotonically toward zero as $\sigma\rightarrow\infty$. In this case, the wave packet tends to expand. For $\Gamma=2\pi$, the variational energy vanishes for any value of $\sigma$. This reflects the scale invariance of the 2D attractive cubic nonlinear Schrodinger equation. For $\Gamma>2\pi$, the energy is negative and decreases without bound as $\sigma\rightarrow0$. This corresponds to collapse within the mean-field model.

While the Gaussian ansatz captures the qualitative features of this critical behavior, the critical coupling strength and the corresponding Townes soliton profiles can be exactly determined by seeking a positive, nodeless stationary solution of Eq.~\ref{eq:townes-dimensionless-gpe} using the ansatz~\cite{PRL.13.479, OL.39.3429, PRL.127.023603}
\begin{equation}
	\phi(\rho,\tau)  =  \frac{  \kappa}{\sqrt{2\Gamma}  }  Q(s)e^{i \kappa^2 \tau /2},
	\qquad s=\kappa\rho, (\kappa>0).
	\label{eq:townes-stationary-ansatz}
\end{equation}
Substitution into Eq.~\eqref{eq:townes-dimensionless-gpe} gives the universal equation
\begin{equation}
	Q''(s) + \frac{1}{s}Q'(s) - Q(s) + Q^3(s) = 0,
	\qquad Q'(0)=0,
	\qquad Q(s\rightarrow\infty)=0.
	\label{eq:townes-radial}
\end{equation}
The exact nodeless solution of Townes soliton satisfies
\begin{equation}
	\Gamma_T\equiv\pi\int_0^\infty ds\,sQ^2(s)  =  5.85044825\ldots.
	\label{eq:townes-exact-threshold}
\end{equation}
It can be verified that the energy of the obtained solution of Townes soliton also satisfies $E_T=0$.

Combining $\Gamma=4\pi N|a_{\downarrow\downarrow}|I_z$ and Eq.~\eqref{eq:townes-exact-threshold} gives the critical atom number of the Townes soliton as
\begin{equation}
	N_T=\frac{\Gamma_T}{4\pi|a_{\downarrow\downarrow}|I_z}.
	\label{eq:townes-critical-number-general}
\end{equation}
For a frozen harmonic-oscillator ground state, Eq.~\eqref{eq:Iz} is applicable. Therefore,
\begin{equation}
	N_T=\frac{\Gamma_T}{\sqrt{8\pi}}
	\frac{a_z}{|a_{\downarrow\downarrow}|}
	\simeq1.167\frac{a_z}{|a_{\downarrow\downarrow}|}.
	\label{eq:townes-critical-number-harmonic}
\end{equation}

%%%%%%%%%%%%%%%%% Fig. S7 %%%%%%%%%%%%%%%%%

%%%%%%%%%%%%%%%%% Fig. S7 %%%%%%%%%%%%%%%%%
